\documentclass[journal]{IEEEtran}
\usepackage{amsmath,amsfonts}
\usepackage{algorithm}
\usepackage{array}
\usepackage[caption=false,font=footnotesize,labelfont=rm,textfont=rm]{subfig}
\usepackage{textcomp}
\usepackage{stfloats}
\usepackage{url}
\usepackage{verbatim}
\usepackage{graphicx}
\usepackage{cite}
\usepackage{algorithmicx}
\usepackage{amsmath}
\usepackage{graphics} 
\usepackage{epsfig} 
\usepackage{mathptmx} 
\usepackage{times} 
\usepackage{amssymb}  
\usepackage{algorithm}
\usepackage{algpseudocode}
\usepackage{indentfirst}
\usepackage{subfloat}
\usepackage{color}
\usepackage{booktabs}
\usepackage{bm}
\usepackage{amsthm}
\newtheorem{theorem}{Theorem}
\newtheorem{lemma}{Lemma}
\newtheorem{definition}{Definition}
\usepackage{multirow}

\begin{document}

\title{EASpace: Enhanced Action Space for  \\ Policy Transfer}
\author{Zheng Zhang, Qingrui Zhang, Bo Zhu, Xiaohan Wang, and Tianjiang Hu
\thanks{All authors are  with Machine Intelligence and Collective Robotics (MICRO) Lab, Sun Yat-sen University, Shenzhen 518107, China (Corresponding author: Qingrui Zhang, {\tt\small zhangqr9@mail.sysu.edu.cn;)}}}



\maketitle

\begin{abstract}

 Formulating expert policies as macro actions promises to alleviate the long-horizon issue via structured exploration and efficient credit assignment. However, traditional option-based multi-policy transfer methods suffer from inefficient exploration of macro action's length and insufficient exploitation of useful long-duration macro actions. In this paper, a novel algorithm named \emph{EASpace} (Enhanced Action Space) is proposed, which formulates macro actions in an alternative form to accelerate the learning process using multiple available sub-optimal expert policies. Specifically, EASpace formulates each expert policy into multiple macro actions with different execution {times}. All the macro actions are then integrated into the primitive action space directly. An intrinsic reward, which is proportional to the execution time of macro actions, is introduced to encourage the exploitation of useful macro actions. The corresponding learning rule that is similar to Intra-option {Q-learning} is employed to improve the data efficiency. Theoretical analysis is presented to show the convergence of the proposed learning rule. The efficiency of EASpace is illustrated by a grid-based game and a multi-agent pursuit problem. The proposed algorithm is also implemented in physical systems to validate its effectiveness.
\end{abstract}

\section{Introduction}
To facilitate the learning process of reinforcement learning (RL) in long-horizon tasks, numerous transfer learning (TL) methods have been extensively studied to convey the knowledge from source tasks to the target task \cite{zhuang2020comprehensive}. One major direction of TL is policy transfer, where the external knowledge takes the form of expert policies derived from source tasks \cite{zhu2020transfer}. When multiple sub-optimal expert policies are available, most policy transfer methods attempted to choose the best one for learning. The best expert policy is selected according to the similarity between Markov decision processes (MDPs) of source and target tasks \cite{song2016measuring,svetlik2017automatic,silva2018object} or the expected performance gain of available expert policies \cite{fernandez2006probabilistic, li2018optimal}. However, such single-policy transfer methods are not efficient enough because the target task could not be resolved optimally by any single expert policy. Therefore, several multi-policy transfer methods are proposed to learn from multiple expert policies concurrently. One of the most popular approaches of multi-policy transfer is policy distillation that mixes the information of all expert policies firstly \cite{rusu2015policy,teh2017distral}. For example, Pan \emph{et al.} minimized the cross entropy between all expert policies and a single policy network named multisource transfer network (MTN). The knowledge of MTN is then transferred to the target task via policy mimic and feature regression \cite{pan2018multisource}. Besides, most methods designed for learning from demonstrations, \emph{e.g.} DQfD \cite{hester2018deep}, could be extended to multi-policy transfer problems by sampling state-action pairs from all expert policies \cite{zhu2020transfer}. Since policy distillation and learning from demonstrations methods provide knowledge of interest from diverse expert policies, it is a necessity to distinguish which part is useful, especially when sub-optimal expert policies exist. However, such a distinguishing process is not trivial in long-horizon tasks due to inefficient exploration and difficult credit assignment, posing another challenge that may be as intractable as solving the original problem.

One possible solution is to formulate expert policies as macro actions, then resort to hierarchical reinforcement learning (HRL) methods for learning when to execute which expert policy and how long it lasts \cite{li2019context,yang2020efficient}. Macro actions are temporally extended actions. Instead of being executed for only one timestep as primitive actions, macro actions make decisions according to some closed-loop policies for a period of time \cite{sutton1999between}. In comparison with traditional multi-policy transfer methods, formulating expert policies as macro actions results in a shortened task horizon, which is promising to alleviate the difficult credit assignment. In addition, structured exploration with expert policies also prevents multi-policy transfer methods from determining the usefulness of expert policies at all timesteps, which reduces the computation burdens dramatically.

\begin{figure}
    \centering
    \includegraphics[width=0.48\textwidth]{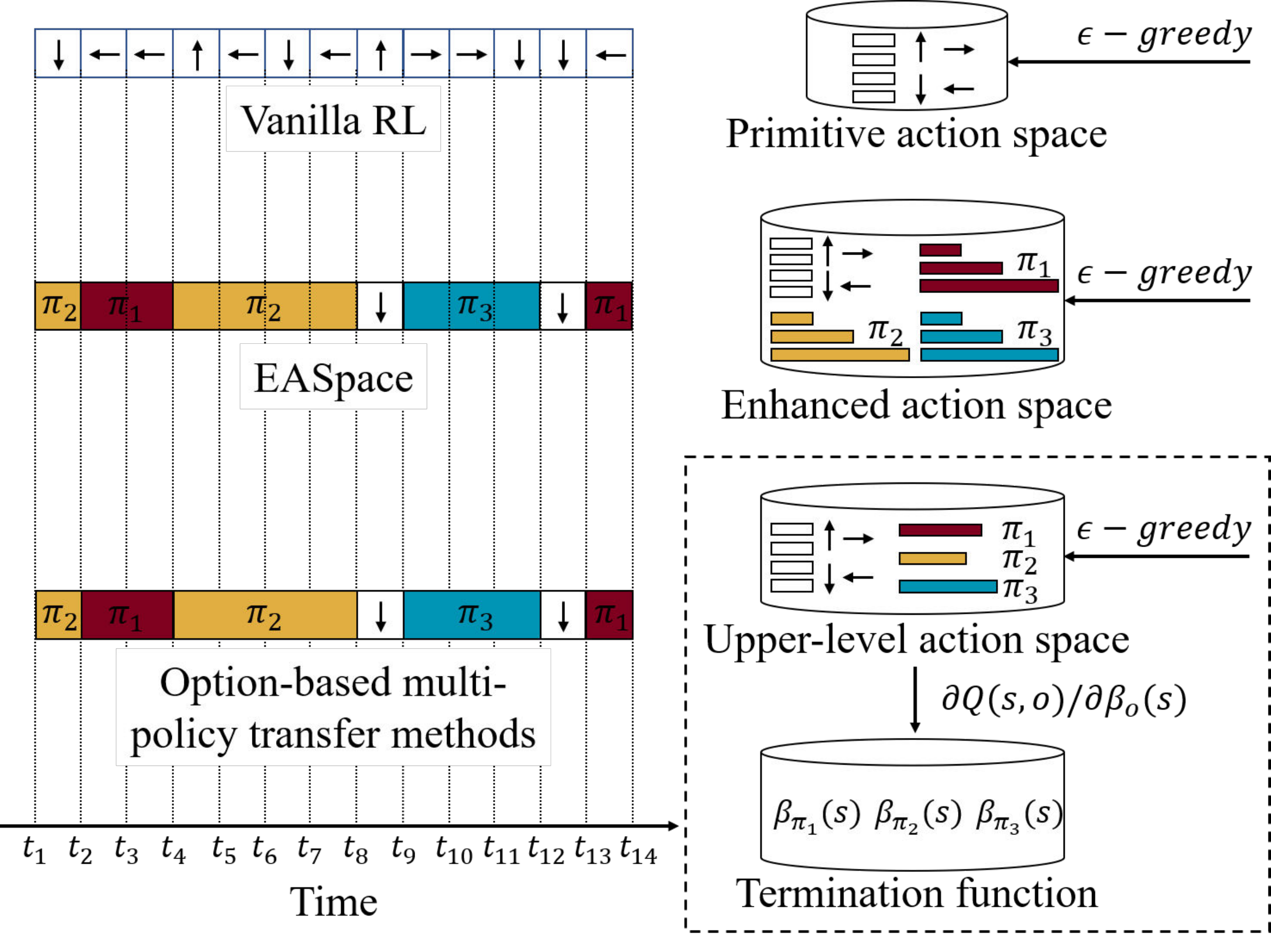}
    \caption{The action sequences generated by different algorithms in one episode. White blocks denote primitive actions and colored blocks denote expert policies. Vanilla RL outputs a primitive action at each timestep, while both EASpace and option-based multi-policy transfer methods formulate expert policies as macro actions, then learn when to execute which expert policy and how long it lasts. The difference between EASpace and option-based methods lies in the structure of action space. EASpace formulates one expert policy as multiple macro actions with different duration, then integrates them into the primitive action space directly. In contrast, there is only one macro action for each expert policy in the option-based action space. The learning of the optimal length depends on the gradient of {the} action value with respect to the termination function.}
    \label{fig:carton}
\end{figure}

The existing macro action-based multi-policy transfer methods are established upon the option-critic framework that simultaneously learns the upper-level policy and the length of macro actions \cite{bacon2017option}. However, the option-critic framework is not efficient enough in the context of multi-policy transfer problems. The first reason is that the learning of the optimal length is inefficient due to the lack of corresponding exploration strategies. In option-based multi-policy transfer methods \cite{li2019context,yang2020efficient}, there is only one macro action for each expert policy, as shown in Fig. \ref{fig:carton}. The length of macro actions is adjusted via the gradient of the action value with respect to the termination function. When the agent follows some exploration strategy, \emph{e.g.} $\epsilon$-greedy, the executed action is selected randomly from the upper-level action space. Although it is possible to try any expert policy, the length of macro actions is determined by the current termination function. It implies that the option-critic framework only exploits the current length of macro actions while never exploring macro actions with different {lengths}. For example, the long-duration macro actions have little chance to be tried before the length of macro actions is extended by updating the parameters of the termination function. Therefore, the learning of the optimal length is doomed to be inefficient due to such an imbalanced exploration-exploitation process. The second reason is that the exploitation of useful long-duration macro actions is insufficient. In theory, primitive actions are sufficient for solving any MDP, which suggests the action values of macro actions are always no more than that of the best primitive action. Although the usefulness of macro actions has been determined, it is still difficult to exploit useful long-duration macro actions due to their lower action values. As a result, the frequent execution of primitive actions makes credit assignment inefficient because the resultant trajectories are still very long. In addition, option-based multi-policy transfer methods suffer from the option-shrink issue, \emph{i.e.} the length of long-duration macro actions always decreases due to their lower values. To tackle this problem, a regularizer is added to the advantage function when updating the parameters of the termination function. However, this approach is very sensitive in that the options tend to degenerate to two trivial solutions: only one active option that solves the whole task, or the agent changes options at every timestep \cite{vezhnevets2017feudal,eysenbach2018diversity,nachum2018data}. Therefore, it is also difficult to employ appropriate macro actions for structured exploration. From the above insights, it can be concluded that the key to macro action-based policy transfer methods is to efficiently learn the usefulness of macro actions and their optimal length, then sufficiently exploit useful long-duration macro actions in the learning process.

In this paper, a multi-policy transfer algorithm called \emph{EASpace} (Enhanced Action Space) is proposed to learn from multiple sub-optimal source tasks by formulating expert policies into macro actions. The objective of EASpace is to learn when to execute macro actions, which macro action should be executed, and how long it lasts. For this objective, each expert policy is formulated into multiple macro actions with different execution {times}. As illustrated in Fig. \ref{fig:carton}, all the macro actions are integrated into the primitive action space directly so that they can be chosen at the same level as primitive actions. Therefore, any standard semi-Markov decision process (SMDP) method can be employed to learn the upper-level policy and the length of macro actions concurrently. In addition, this formulation integrates the exploration of the length of macro actions into the standard RL exploration-exploitation process. Macro actions with various {lengths} are accessible to the agent throughout the training process, which is significant for the learning of the optimal length. Since the action values of macro actions are always no more than that of the best primitive action, an intrinsic reward called \emph{macro action bonus} is introduced to present the preference for the long-duration macro actions. In comparison with option-based methods, the length of macro actions is more easily measured in EASpace due to its special formulation. So the intrinsic reward is defined to be proportional to the duration of macro actions, which encourages the exploitation of long-duration macro actions with good performance. Similar to Intra-option {Q-learning} in option-based methods \cite{sutton1999between}, the \emph{Intra-Macro-Action Learning Rule} (IMALR) customized for EASpace is proposed by adjusting the temporal difference target (TD target). IMALR enables gathering multiple training data at each timestep during the execution of macro actions, which is significant to improve data efficiency. The overall contributions of this paper are summarized as follows.

\begin{enumerate}
    \item A novel policy transfer algorithm called \emph{EASpace} is proposed to transfer the knowledge of multiple sub-optimal expert policies. EASpace formulates each expert policy into multiple macro actions with different execution {times}, which not only enables the concurrent learning of the upper-level policy and the optimal length of macro actions but also provides an appropriate exploration strategy for the learning of the optimal length.
    \item An intrinsic reward, named \emph{macro action bonus}, is designed to encourage the exploitation of long-duration macro actions with good performance, which is significant for all macro action-based methods to explore structurally and ease credit assignment.
    \item The \emph{Intra-Macro-Action Learning Rule} (IMALR) is proposed to improve the data efficiency by adjusting the TD target of macro actions.
    \item {A} rigorous theoretical analysis is conducted to show the convergence of IMALR. 
    \item Real-world experiments are conducted to justify the feasibility of deploying the policies learned from EASpace to real systems.
\end{enumerate}

The rest of this paper is organized as follows. In Section \ref{sec:related_work}, the related works are summarized. Preliminaries are provided in Section \ref{sec:preliminary}. Section \ref{sec:methodology} presents the implementation details and theoretical analysis of the proposed algorithm. Experiments and results are given in Section \ref{sec:experiments_and_results}. Finally, conclusions and future works are available in Section \ref{sec:conclusion}.

\section{Related works}\label{sec:related_work}
\subsection{Policy Transfer in Reinforcement Learning}

Various heuristic metrics are proposed to measure the usefulness of expert policies via the descriptions of MDPs \cite{song2016measuring,silva2018object,svetlik2017automatic}. After selecting the most suitable expert policy, traditional policy transfer methods, \emph{e.g.} model-reference RL \cite{zhang2021model}, are employed to learn from the best expert policy throughout the training process. To transfer the knowledge of multiple expert policies,  the notion of expected performance gain is proposed to record the transfer benefits obtained in the previous training process \cite{fernandez2006probabilistic,li2018optimal}. At the beginning of each episode, the agent probabilistically selects one expert policy to transfer according to the expected performance gain. Since the expected performance gain is updated online, the selected expert policy changes in each episode, which makes the resultant policy embrace the knowledge from multiple policies. To learn from multiple expert policies simultaneously in one episode, policy distillation is employed to mix the knowledge of expert policies via minimizing the MSE loss \cite{rusu2015policy} between the mixed policy and expert policies or the KL divergence \cite{teh2017distral,yin2017knowledge}. The integrated knowledge is used to initialize the weights of the action value function \cite{parisotto2015actor} or add a supervised term to the RL objective function \cite{pan2018multisource}. Different from mixing multiple expert policies firstly, Barreto \emph{et al.} combined the techniques of successor features and generalized policy improvement theorem, providing performance guarantees for the transferred policy \cite{barreto2017successor}. Besides the above algorithms, most learning from demonstrations methods can be used to transfer the knowledge of multiple expert policies simultaneously \cite{zhu2020transfer}. These methods mix the demonstrations sampled from all expert policies firstly. The demonstrations are then employed to twist the objective function \cite{bertsekas2011approximate,piot2014boosted,chemali2015direct}, shape the reward function \cite{brys2015reinforcement}, fill the replay buffer \cite{vecerik2017leveraging,hester2018deep}, or bias the sampling strategy \cite{sui2020formation,yan2020towards,yan2021deep}. But as mentioned above, determining when to learn from which expert policy at all timesteps is time-consuming for most multi-policy transfer methods, which may result in negative transfer.

\subsection{Hierarchical Reinforcement Learning}
 One of the major directions of hierarchical RL is the option framework, in which the macro actions are regulated by three components, \emph{i.e.} initial state set,  lower-level policy, and termination function \cite{sutton1999between}. Based on this formulation, Intra-option {Q-learning} algorithm is proposed to extract more data during the execution of macro actions, which results in greater data efficiency than traditional SMDP methods. However, the termination function and the lower-level policy are usually pre-trained or predefined in Intra-option {Q-learning}. To tackle this problem, Bacon \emph{et al.} proposed the option-critic architecture that enables the concurrent learning of all components of the options together with the upper-level policy \cite{bacon2017option}. Then, an entropy regularizer and a mutual information term are introduced into the learning process of the lower-level policy to encourage the diversity and individuality of macro actions \cite{zhu2021empowering}. Since the option-critic architecture is highly customized to the option-based SMDP, Zhang \emph{et al.} proposed DAC that reformulates the option framework as two parallel {augmented} MDPs. It enables the hierarchy to be learned by most off-the-shelf RL optimization methods \cite{zhang2019dac}. However, both the option-critic architecture and DAC suffer from the non-stationarity issue derived from the changing lower-level policy. Yang \emph{et al.} alleviated this problem by adaptively limiting the exploration of the lower-level agent, but the impact of changing parameters of the lower-level policy remains to hinder the learning \cite{yang2021hierarchical}. In the context of multi-policy transfer, \cite{li2019context,yang2020efficient} formulated the expert policies as options, then employed option-critic architecture to learn the upper-level policy and the termination function concurrently. However, the option-critic architecture is not efficient enough in multi-policy transfer problems due to inefficient exploration of macro action's {lengths} and insufficient exploitation of useful long-duration options.

Besides the option framework, Feudal RL is another mainstream in hierarchical RL \cite{dayan1992feudal}. {In Feudal RL, the upper-level policy sets a goal for the lower-level policy, after which the lower-level policy learns to accomplish the goal by maximizing an intrinsic reward function \cite{dilokthanakul2019feature, zhou2022online,ZHOU2019443}}. The execution of the lower-level policy terminates once the goal is reached, then one training data is extracted to train the upper-level policy via SMDP \cite{kulkarni2016hierarchical}. Since the hierarchy is learned concurrently, the non-stationarity issue is also inevitable due to the changing lower-level policy. Hence, the transition relabeling techniques are proposed to stabilize the training process \cite{levy2017learning,nachum2018data}. Most existing Feudal RL methods manually design the goal space \cite{kulkarni2016hierarchical,tang2018hierarchical} or use states as goals directly \cite{nachum2018data,levy2017learning}, which makes the algorithm less task-agnostic or leads to an overlarge goal space. To alleviate this issue, Pateria \emph{et al.} proposed to make up the goal space only with the states that are frequently encountered in the training process \cite{pateria2021end}. Besides, learning a low-dimensional goal space is also possible to reduce the search space of the higher-level policy \cite{vezhnevets2017feudal}. However, since most Feudal RL methods parameterize the lower-level policy with the universal value function approximator \cite{schaul2015universal}, it is hard to characterize multiple expert policies with one universal policy. To our best knowledge, there is no multi-policy transfer method using the feudal RL architecture.

\section{Preliminaries}\label{sec:preliminary}
\subsection{Reinforcement Learning}
 A Markov Decision Process can be described by a tuple $(\mathcal{S},\mathcal{A},\mathcal{P},\mathcal{R},\gamma)$, where $\mathcal{S}$ is the state space, $\mathcal{A}$ is the action space, $\mathcal{P}$ is the state-transition model, $\mathcal{R}$ is the reward function, $\gamma$ is the discount factor. The objective of RL is to learn an optimal policy $\pi^*(a|s)$ that maximizes the expectation of accumulated rewards $\mathbb{E}[R_t]=\mathbb{E}[\sum_{k=t}^T\gamma^{k-t} r_t]$, where $s$ is the environment state, $a$ is the executed action and $r_t$ is the reward obtained at the timestep $t$, $\pi(a|s)$ is the probability of choosing the action $a$ at the state $s$. The state-value function $V^\pi(s)$ is the expected accumulated rewards conditioned on a given state,
 {
 \begin{equation}
     V^{\pi}(s)=\sum_{a}\pi(a|s)\sum_{s'}\mathcal{P}(s'|s,a)\big(r+\gamma V^{\pi}(s') \big), 
 \end{equation}}
 where $s'$ is the state at the next timestep. The action value function is introduced to describe the usefulness of a state-action pair, which is defined as
  {
 \begin{equation}
     Q^{\pi}(s,a)=r+\gamma \mathbb{E}_{s'}[V^{\pi}(s')].
 \end{equation}}
 Note that both state and action value functions depend on the policy $\pi$. When the policy is optimal, the corresponding action-value function is called the optimal action-value function. It is the maximal expected accumulated rewards achievable by following any policy conditioned on a given state-action pair,
  {
 \begin{equation}
  Q^*(s,a)=\max_\pi\mathbb{E}[R_t|s,a,\pi]=\mathbb{E}[r+{\gamma}\max_{a}Q^{*}(s',a)]. 
 \end{equation}}
 The aforementioned equation is the famous Bellman equation that can be solved by an iterative style in {Q-learning} \cite{silver2016mastering}.
  {
 \begin{equation}
     Q_{i+1}(s,a)=Q_{i}(s,a)+\alpha[r+{\gamma}\max_{a}Q_{i}(s',a)-Q_{i}(s,a)],
 \end{equation}}
where $\alpha$ is the learning rate and $r+{\gamma}\max_{a}Q_{i}(s',a)$ is the TD target.
 
\subsection{Semi-Markov Decision Process}
Rather than choosing an action at each timestep as in MDP, in SMDP actions may last multiple timesteps, which is the situation where the action space is composed of macro actions. The optimal action value function in SMDP is 
 {
\begin{equation}
 Q^*(s_t,a_t)=\mathbb{E}[r_{t}+\cdots+\gamma^{k-1}r_{t+k-1}+{\gamma}^k\max_{a}Q^{*}(s_{t+k},a)],  
\end{equation}
}
 where $a_t$ is initiated at timestep $t$ and terminates at timestep $t+k-1$. Hence, SMDP {Q-learning} \cite{bradtke1994reinforcement} updates the action value after the termination of each macro action by 
 \begin{equation} \label{equ:SMDP_update}
 \begin{split}
  Q_{i+1}(s_t,a_t)=&Q_{i}(s_t,a_t)+\alpha[r_{t}+\cdots+\gamma^{k-1}r_{t+k-1}\\
  &+{\gamma}^k\max_{a}Q_{i}(s_{t+k},a)-Q_{i}(s_t,a_t)],
 \end{split}
 \end{equation}
 where $r_{t}+\gamma r_{t+2}+\cdots+\gamma^{k-1}r_{t+k-1}+{\gamma}^k\max_{a}Q_{i}(s_{t+k},a)$ is the TD target. Since SMDP {Q-learning} treats macro actions as opaque indivisible units, it only gathers one training data during the execution of macro actions. Based on the option framework \cite{sutton1999between}, Sutton \emph{et al.} proposed Intra-option {Q-learning} to extract multiple transitions at every timestep by rewriting the TD target as $(1-\beta(s',a))Q(s',a)+\beta(s',a)\max_{a}Q(s',a)$, where $a$ is the action executed at this timestep, $s'$ is the state at the next timestep, $\beta(s',a) $ is the probability of $a$ terminating at the next state $s'$.
\section{Methodology}\label{sec:methodology}
\begin{figure*}
    \centering
    \includegraphics[width=0.95\textwidth]{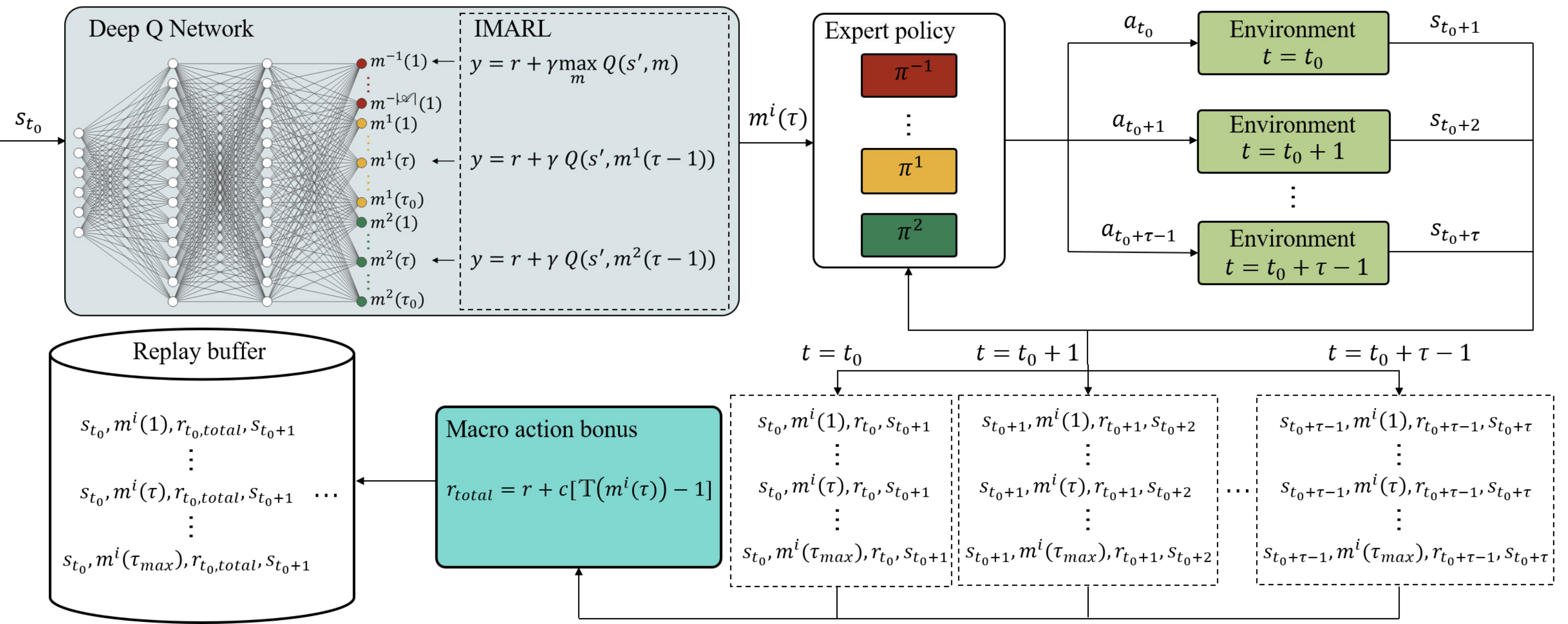}
    \caption{The overview of EASpace. Taking two expert policies for example, the deep Q network outputs action values for both primitive actions (red nodes) and macro actions (yellow and green nodes). The macro actions with the same color are derived from the same expert policy, and their duration ranges from one timestep to $\tau_0$ timesteps. IMARL is proposed to extract multiple transitions at each timestep by adjusting the TD target $y$. Before storing transitions into the replay buffer, the task reward is combined with an intrinsic reward named macro action bonus.}
    \label{fig:EASpace}
\end{figure*}

In this paper, the multi-policy transfer problem is defined as follows: given a set of expert policies $\Pi={\pi^1,\pi^2,\cdots,\pi^n}$, the objective is to utilize the knowledge of $\Pi$ to quickly learn an optimal policy for the target task. Although each expert policy gives advice, \emph{i.e.} an action $a=\pi^i(s)$, all the time, they are far from optimal. It is possible that some expert policies give totally wrong advice in some states, so the agent needs to learn which expert policy is more valuable in a given state. There is also no guarantee that helpful expert policies always exist, hence it is requisite for the agent to decide whether {to learn} from expert policies or {explore} the environment by itself. The whole structure of EASpace is illustrated in Fig. \ref{fig:EASpace} and Algorithm \ref{alg:EASpace}. We will describe the proposed algorithm from three aspects, \emph{i.e.} enhanced action space via macro actions, macro action bonus, and intra-macro-action learning rule.

\begin{algorithm}
    \caption{EASpace}\label{alg:EASpace}
    \renewcommand{\algorithmicrequire}{\textbf{Input:}}
    \renewcommand{\algorithmicensure}{\textbf{Output:}}
    \begin{algorithmic}[1]
		\Require Policy update frequency $T_u$, target network update frequency $f$, replay memory $\mathcal{D}$, action-value network $Q$ with random weights $\theta$, target network $Q^-$ with weights $\theta ^-=\theta$
		\Ensure Optimal policy $Q$
        \For{$episode=1,2,...,$}
            \State Initialize environment state $s$
            \State $\tau_r=1$
            \For{$t=1,2,..,$}
                \State $\tau_r=\tau_r-1$
                \If {$\tau_r=0$}
                    \State Select an upper-level action $m^i(\tau)$ by $\epsilon$-greedy
                    \State The remaining execution time $\tau_r=\tau$
                \EndIf
                \State Execute the lower-level action $a$ according to $\pi^i$
                \State Observe the reward $r$ and new observation $s'$
                \For{$\tau_j=1,2,...,\tau_{max}$}
                    \State $r_{total}=r+c[{\mathbb{T}}(m^i(\tau_j))-1]$
                    \State Store the transition $(s,m^i(\tau_j),r_{total},s')$ in the replay buffer $\mathcal{D}$
                \EndFor
            \EndFor
            \For{$update=1,2,...T_u$}
                \State Sample transitions from $\mathcal{D}$
                \State Learn according to IMALR (Algorithm \ref{alg:Intra-macro-action_Q_Learning})
                \State Every $f$ steps copy $\theta$ to $\theta^-$
            \EndFor
        \EndFor
    \end{algorithmic}
\end{algorithm}

\subsection{Enhanced Action Space with Macro Actions}

Traditional multi-policy transfer methods mix all expert policies via policy distillation \cite{rusu2015policy,teh2017distral,pan2018multisource} or demonstrations \cite{brys2015reinforcement,wiewiora2003principled}. The integrated information is then transferred to the target task. Hence, it is requisite to learn which part of the integrated information is more informative at every timestep, leading to heavy computation burdens. However, in multi-policy transfer problems, one expert policy with good performance in a certain state is likely to be useful in nearby states, while a totally useless expert policy at one timestep seems also uninformative in several following timesteps. This intuition implies the usefulness of expert policies is not {necessarily} to be determined at all timesteps. Therefore, EASpace resorts to macro actions for structured exploration and efficient credit assignment. Specifically, EASpace formulates each expert policy $\pi^i$ into $\tau_0$ macro actions $m^i(1),m^i(2),\cdots,m^i(\tau_0)$, where the numbers in brackets indicate the duration (\emph{i.e.} the execution timesteps) of macro actions. The superscript $i$ implies this macro action is derived from the expert policy $\pi^i$. $\tau_0$ is a hyperparameter that indicates the maximal duration of macro actions. After that, EASpace integrates all $n \times \tau_0$ macro actions into the primitive action space directly, where $n$ is the number of expert policies and $\tau_0$ is the number of macro actions created for each expert policy. This special formulation transforms the learning of the optimal length of $\pi^i$ into the problem of learning the optimal macro action among $m^i(1),m^i(2),\cdots,m^i(\tau_0)$.  Therefore, the concurrent learning of the upper-level policy together with the optimal length can be accomplished by any standard SMDP method via finding out the optimal action from the integrated action space.  Note that primitive actions can be treated as 1-timestep macro actions that follow state-irrespective expert policies. So we denote primitive actions as $m^{-1}(1),\cdots, m^{-|\mathcal{A}|}(1)$. Under this notation, the integrated action space can be represented as  $$\{m^i(\tau)|i,\tau \in \mathbb{Z},-|\mathcal{A}| \leq i \leq n,i \neq 0,1\leq \tau \leq \tau_{max}\},$$ where $\tau_{max}=1$  if $-|\mathcal{A}| \leq i \leq -1$ and $\tau_{max}=\tau_0$ if $1 \leq i \leq n$.

As illustrated in Fig. \ref{fig:EASpace}, a DQN-based algorithm is employed in this paper. After observing the environment state $s_{t_0}$ at the timestep $t_0$, the deep Q network outputs action values for all actions in the integrated action space. Once one higher-level action $m^i(\tau)$ is selected, the agent executes lower-level actions according to the corresponding policy $\pi^i$ and the current state for $\tau$ timesteps, $a_{t_0}=\pi^i(s_{t_o}),\cdots,a_{t_0+\tau-1}=\pi^i(s_{t_0+\tau-1})$. Before the timestep $t_0+\tau$, there is no other action selection process for the deep Q network. Note that all macro actions ranging from one timestep to $\tau_0$ timesteps are accessible in the above action selection process. It implies the integrated action space provides an appropriate exploration strategy for the learning of the optimal length of macro actions.  For example, rather than extend a useful macro action firstly as option-based methods do, EASpace is able to try its long-duration counterpart at the very beginning of the training process. Therefore, any standard exploration strategy, \emph{e.g.} $\epsilon$-greedy, can be employed to balance the exploitation of the optimal length with the exploration of different {lengths}, which results in more efficient concurrent learning of the upper-level policy and the optimal length.

\subsection{Macro Action Bonus}
Through the investigation of the influence of different macro actions, McGovern \emph{et al.} found that good macro actions accelerate the learning process while inappropriate ones may result in negative transfer \cite{mcgovern1997roles}. One reason is that the structured exploration with good macro actions leads to more successful experiences while bad macro actions always keep the agent away from the task goal. In addition, in order to shorten the task horizon and ease credit assignment via temporal abstraction, long-duration macro actions should be executed more frequently than short-duration macro actions. Therefore, the key to macro action-based multi-policy transfer methods is sufficient exploitation of both useful and long-duration macro actions. However, only integrating macro actions into the primitive action space can not achieve it. We will explain the reason in the context of $\epsilon$-greedy exploration strategy.

At the $\epsilon$-stage, the probability of being chosen is equal for all actions in the integrated action space. It implies that useful macro actions are executed as frequently as useless ones. At the greedy-stage, the agent selects the action with maximal action value. However, if the action value function is well approximated, the action value of $m^i(\tau)$ at timestep $t_0$ is always no more than $m^i(1)$ since the policy $\pi^i$ may be sub-optimal for $s_{t_0+1},\cdots,s_{t_0+\tau-1}$. Therefore, the long-duration macro actions with good enough performance are also not selected even if the agent has learned which expert policy is useful.

To tackle this problem, an intrinsic reward named \emph{macro action bonus} is introduced to present the preference for long-duration macro actions. This intrinsic reward, which is proportional to the duration of macro actions, will be combined with the task reward as shown in Line 13 of Algorithm \ref{alg:EASpace}, where the operator $\mathbb{T}(\cdot)$ returns the duration of macro actions and $c$ is a positive number. If a long-duration macro action is helpful to the target task, \emph{i.e.} its action value is equal or slightly less than that of the optimal action, the agent will exploit it frequently at the greedy-stage due to the additional intrinsic reward. On the contrary, if one macro action is totally useless, the intrinsic reward will not lead to a dramatic increase in its action value, so uninformative macro actions remain abandoned. Therefore, the macro action bonus enables sufficient exploitation of useful long-duration macro actions, which is significant to employ temporal abstraction for structured exploration and efficient credit assignment.  In comparison, option-based methods add a regularizer to the advantage function to avoid the option-shrink issue. However, the action values of macro actions are still no more than that of the best primitive action. Since the termination probability at future states is not available, the length of macro actions is not measurable in option-based methods. So it is also difficult to twist the action value of macro actions via an intrinsic reward as EASpace does. Consequently, insufficient exploitation of long-duration macro actions makes the algorithm degenerate to vanilla RL methods to some extent.

\subsection{Intra-Macro-Action Learning Rule (IMALR)}
As mentioned above, any SMDP method can be employed to learn the upper-level policy and the optimal length of macro actions after integrating macro actions into the primitive action space.  However, Sutton \emph{et al.} pointed out that the SMDP methods are not efficient enough since only one training data is extracted during the execution of macro actions. To acquire more training data, they proposed Intra-option {Q-learning} that employs the Markov property of options and special temporal difference methods \cite{sutton1999between}. In this paper, the expert policies are also Markov, so the lower-level actions suggested by $\pi^i$ only depend on the state, $a_t=\pi(s_t)$. Due to this property, the transition $(s_t,m^i(\tau),r_t,s_{t+1})$, which is valid for $m^i(\tau)$, is also valid for other macro actions derived from $\pi^i$, so we can collect multiple transitions at each timestep as \cite{sutton1999between} does (Line 12-15 in Algorithm \ref{alg:EASpace}). Note that the transition $(s_t,m^i(\tau),r_t,s_{t+1})$ means the state at timestep $t$ is $s_t$ during the execution of $m^i(\tau)$ with $r_t$  as the immediate reward and  $s_{t+1}$ the state at timestep $t+1$. It implies that  $Q(s_{t},m^i(\tau))$ can not be learned as (\ref{equ:SMDP_update}) since one transition does not embrace the information of the accumulated reward. Motivated by the aforementioned observations, IMALR is proposed to learn the action value of macro actions via temporal difference.

Consider the action value of a multi-timestep macro action $m^i(\tau)$ initiated at timestep $t_0$,
\begin{equation} \label{equ:action_value_tau}
   Q(s_{t_0},m^i(\tau))=\mathbb{E}[r_{t_0}+\cdots+r_{t_0+\tau-1}+\max_m Q(s_{t_0+\tau},m)],
\end{equation}
where $\tau>1$, $r_{t_0},\cdots,r_{t_0+\tau-1}$ are rewards received following the policy $\pi^i$. $\max_m Q(s_{t_0+\tau},m)$ is the maximal action value over the integrated action space. This equation implies that after executing $m^i(\tau)$ for $\tau$ timesteps the agent has the chance to select an optimal action in the integrated action space.  Note that the discount factor $\gamma$ is neglected for clarity. Similarly, the action value of $m^i(\tau-1)$ that is derived from the same policy $\pi^i$ but initiated at timestep $t_0+1$ is
\begin{equation} \label{equ:action_value_tau-1}
   Q(s_{t_0+1},m^i(\tau-1))=\mathbb{E}[r_{t_0+1}+\cdots+r_{t_0+\tau-1}+\max_m Q(s_{t_0+\tau},m)],
\end{equation}
where $r_{t_0+1},\cdots,r_{t_0+\tau-1}$ are rewards following the same policy $\pi^i$.
From equations (\ref{equ:action_value_tau}) and (\ref{equ:action_value_tau-1}), it is obvious that we have
\begin{equation} \label{equ:td_target_for_macro_action}
    Q(s_{t_0},m^i(\tau))=\mathbb{E}[r_{t_0}+\gamma Q(s_{t_0+1},m^i(\tau-1))].
\end{equation}
The intuition behind (\ref{equ:td_target_for_macro_action}) is that, no matter whether executing $m^i(\tau)$ from timestep $t_0$ or executing $m^i(\tau-1)$ from timestep $t_0+1$, the resultant experiences are the same because both of them follow the same policy until the same timestep.

The action value of one-step macro actions that are initiated at timestep $t_0$ is the same as vanilla {Q-learning},
\begin{equation} \label{equ:td_target_for_primitive_action}
    Q(s_{t_0},m^i(1))=\mathbb{E}[r_{t_0}+\gamma \max_{m} Q(s_{t_0+1},m)].
\end{equation}
The reason is that in both scenarios actions only last for one timestep. At the next timestep, the agent can choose an optimal action again.

The above Bellman equations (\ref{equ:td_target_for_macro_action}) and (\ref{equ:td_target_for_primitive_action}) suggest a special TD target for $m^i(\tau)$, 
\begin{equation} \label{equ:td_target}
    y=\begin{cases}
    r+\gamma \max_{m} Q(s',m), & \text{if } \tau=1, \\
    r+\gamma Q(s',m^i(\tau-1)), & \text{if } \tau>1.
    \end{cases}
\end{equation}
Note that the TD target in (\ref{equ:td_target}) only involves the reward at the current timestep, enabling EASpace to learn from transitions that do not contain the information of accumulated reward. The whole IMALR is described in Algorithm \ref{alg:Intra-macro-action_Q_Learning}.

 \begin{algorithm}
    \caption{IMALR}\label{alg:Intra-macro-action_Q_Learning}
    \begin{algorithmic}[1]
		\Require Transition $(s,m^i(\tau),r,s')$, action value network $Q$, target network $Q^-$
        \If{$\tau>1$}
            \State TD target $y=r+\gamma Q(s',m^i(\tau-1);\theta^-)$
        \Else
            \State TD target $y=r+\gamma \max_{m} Q(s',m;\theta^-)$
        \EndIf
        \State Calculate loss $L=\frac{1}{2}\big(y-Q(s,m^i(\tau);\theta)\big)^2$
        \State Update the network $Q$ by stochastic gradient descent
    \end{algorithmic}
\end{algorithm}

\subsection{Theoretical Analysis}
Since IMALR extracts transitions at each timestep and employs a customized TD target in comparison with traditional SMDP methods, the convergence property of EASpace is to be discussed. In this section, the theoretical analysis is provided to prove that EASpace converges to the optimal action value with probability 1. Our proof mainly employs the contraction property of the action value iteration operator and the techniques of stochastic approximation. We begin with the definition of the action value iteration operator in EASpace.

\begin{definition}
The action value iteration operator in EASpace is defined as
\begin{equation} 
\begin{aligned}
    (\boldsymbol{H}Q)&(s,m^i(\tau))= \\
    &\begin{cases}
    \sum_{s'}P(s'|s,a)[r+\gamma \max_m Q(s',m)], &\text{if } \tau=1, \\
    \sum_{s'}P(s'|s,a)[r+\gamma Q(s',m^i(\tau-1))], &\text{if } \tau>1,
    \end{cases}
\end{aligned}
\end{equation}
where $a$ is the lower-level action that is executed following the policy $\pi^i$. 
\end{definition}

The action value iteration operator $\boldsymbol{H}$  maps an action value function $Q_1$ to another action value function $Q_2$, $Q_2=\boldsymbol{H}Q_1$. Note that the optimal action value function $Q^*$ is the fixed point of $\boldsymbol{H}$, \emph{i.e.} $Q^*=\boldsymbol{H}Q^*$, because the optimal policy always selects the action with maximal action value after the execution of macro actions.

\begin{lemma} \label{lemma:1}
The action value iteration operator of EASpace is contractive with respect to the infinite norm, which means for any two action value functions $Q_j$ and $Q_k$ we have $\|\boldsymbol{H}Q_j-\boldsymbol{H}Q_k\|_\infty \leq \gamma \|Q_j-Q_k \|_\infty$.
\end{lemma}

\begin{proof}
For any state-action pair $(s,m^i(\tau))$ where $\tau=1$, we have
 {
\begin{equation}
\begin{split}
    &\lvert(\boldsymbol{H}Q_j)(s,m^i(\tau))-(\boldsymbol{H}Q_k)(s,m^i(\tau))\rvert\\
    =&\lvert\sum_{s'}P(s'|s,a)[r+\gamma \max_{m}Q_j(s',m)-r-\gamma \max_{m}Q_k(s',m)]\rvert\\
    =&\lvert\sum_{s'}\gamma P(s'|s,a)[\max_{m}Q_j(s',m)-\max_{m}Q_k(s',m)]\rvert\\
    \leq& \sum_{s'}\gamma P(s'|s,a) \lvert \max_{m}Q_j(s',m)-\max_{m}Q_k(s',m) \rvert\\
    \leq& \sum_{s'}\gamma P(s'|s,a) \max_{m} \lvert Q_j(s',m)-Q_k(s',m) \rvert\\
    \leq& \sum_{s'}\gamma P(s'|s,a) \max_{s,m} \lvert Q_j(s,m)-Q_k(s,m) \rvert\\
    =& \gamma \max_{s,m} \lvert Q_j(s,m)-Q_k(s,m) \rvert\\
    =& \gamma \|Q_j-Q_k\|_\infty.
\end{split}
\end{equation}
}

For any state-action pair $(s,m^i(\tau))$ where $\tau>1$, we have
 {
\begin{equation}
\begin{split}
    &\lvert(\boldsymbol{H}Q_j)(s,m^i(\tau))-(\boldsymbol{H}Q_k)(s,m^i(\tau))\rvert\\
    =&\lvert\sum_{s'}\gamma P(s'|s,a)[Q_j(s',m(\tau-1))-Q_k(s',m(\tau-1))]\rvert\\
    \leq& \sum_{s'}\gamma P(s'|s,a) \lvert Q_j(s',m(\tau-1))-Q_k(s',m(\tau-1)) \rvert\\
    \leq& \sum_{s'}\gamma P(s'|s,a) \max_{m} \lvert Q_j(s',m)-Q_k(s',m) \rvert\\
    \leq& \sum_{s'}\gamma P(s'|s,a) \max_{s,m} \lvert Q_j(s,m)-Q_k(s,m) \rvert\\
    =& \gamma \max_{s,m} \lvert Q_j(s,m)-Q_k(s,m) \rvert\\
    =& \gamma \|Q_j-Q_k\|_\infty.
\end{split}
\end{equation}
}

Therefore, for any action $m^i(\tau)$ in the integrated action space we have $\lvert (\boldsymbol{H}Q_j)(s,m^i(\tau))-(\boldsymbol{H}Q_k)(s,m^i(\tau)) \rvert \leq \gamma \|Q_j-Q_k\|_\infty$, which suggests $\|\boldsymbol{H}Q_j-\boldsymbol{H}Q_k\|_\infty \leq \gamma \|Q_j-Q_k\|_\infty$.
\end{proof}

To prove the convergence of EASpace, some techniques of stochastic approximation are necessary. The following lemma is introduced from \cite{jaakkola1993convergence}.
\begin{lemma} \label{lemma:2}
A random iterative process $\Delta_{n+1}(x)=(1-\alpha_n(x))\Delta_n(x)+\beta_n(x)F_n(x)$ converges to zero with probability 1 under the following assumptions:
\begin{enumerate}
    \item The state space is finite.
    \item $\sum_n \alpha_n(x)=\infty$, $\sum_n \alpha_n^2(x)<\infty$, $\sum_n \beta_n(x)=\infty$, $\sum_n \beta_n^2(x)<\infty$, and $\mathbb{E}[\beta_n|P_n] \leq \mathbb{E}[\alpha_n|P_n]$ uniformly with probability 1.
    \item $\| \mathbb{E}[F_n|P_n] \|_W \leq \gamma \|\Delta_n\|_W$, where $\gamma \in (0,1)$.
    \item Var$[F_n(x)|P_n] \leq C(1+ \|\Delta_n\|_W)^2$, where $C$ is some constant.
\end{enumerate}
Here $P_n=\{\Delta_n,\Delta_{n-1},\cdots,F_{n-1},\cdots,\alpha_{n-1},\cdots,\beta_{n-1},\cdots\}$ stands for the past at step $n$. $F_n(x)$, $\alpha_n(x)$ and $\beta_n(x)$ are allowed to depend on the past insofar as the above conditions remain valid. The notation $\|\cdot\|_W$ refers to some weighted maximum norm. 
\end{lemma}

\begin{proof}
Proof details are given in \cite{jaakkola1993convergence}.
\end{proof}

Now, we give the convergence analysis for EASpace.
\begin{theorem}
The EASpace algorithm converges to the optimal action value function $Q^*(s,m^i(\tau))$ with probability 1 if
\begin{enumerate}
\item The state and action spaces are finite.
\item $\sum_t \alpha_t=\infty$ and $\sum_t \alpha_t^2<\infty$ uniformly with probability 1, where $\alpha$ is the learning rate. 
\item The variance of reward Var$[r]$ is bounded.
\item The discount factor $\gamma \in (0,1)$.
\end{enumerate}
\end{theorem}

\begin{proof}
If $\tau=1$, the learning rule is
 {
\begin{equation}
    Q_{t+1}(s,m^i(\tau))=(1-\alpha)Q_t(s,m^i(\tau))+\alpha(r+\gamma \max_{m}Q_t(s',m)).
\end{equation} 
}
Define 
 {
\begin{equation}
    \Delta_t(s,m^i(\tau))=Q_t(s,m^i(\tau))-Q^*(s,m^i(\tau)),
\end{equation}
}
 {
\begin{equation}
    F_t(s,m^i(\tau))=r+\gamma \max_{m}Q_t(s',m)-Q^*(s,m^i(\tau)).
\end{equation}
}
By subtracting $Q^*(s,m^i(\tau))$ from both sides of the learning rule, the resultant equation has the form of the process in Lemma \ref{lemma:2}. To prove that condition 3 in Lemma \ref{lemma:2} is satisfied for one-step macro action, we have
 {
\begin{equation}
\begin{split}
    &\mathbb{E}[F_t(s,m^i(\tau))|P_n]\\
    =&\sum_{s'}P(s'|s,a)[r+\gamma \max_{m}Q_t(s',m)-Q^*(s,m^i(\tau))]\\
    =&(\boldsymbol{H}Q_t)(s,m^i(\tau))-Q^*(s,m^i(\tau))\\
    =&(\boldsymbol{H}Q_t)(s,m^i(\tau))-(\boldsymbol{H}Q^*)(s,m^i(\tau)).
\end{split}
\end{equation}
}
From Lemma \ref{lemma:1} we have 
\begin{equation} \label{equ:exp_onestep}
    \mathbb{E}[F_t(s,m^i(\tau))|P_n] \leq \gamma \|Q_t-Q^*\|_\infty=\gamma \|\Delta_t\|_\infty.
\end{equation}
The variance of $F_t(s,m^i(\tau))$ given $P_n$ is 
\begin{equation}
\begin{split}
    &{\rm Var}[F_t(s,m^i(\tau))|P_n] \\
    =&{\rm Var}(r+\gamma \max_{m}Q_t(s',m)-Q^*(s,m^i(\tau))|P_n)\\
    =&{\rm Var}(r+\gamma \max_{m}Q_t(s',m)|P_n).
\end{split}
\end{equation}
Since the variance of reward is bounded, it is obvious that there exists a constant $C$ that satisfies
\begin{equation} \label{equ:var_onestep}
    {\rm Var}[F_t(s,m^i(\tau))|P_n] \leq C(1+\| \Delta_t \| ^2_W).
\end{equation}

If $\tau>1$, the learning rule is 
 {
\begin{equation}
    Q_{t+1}(s,m^i(\tau))=(1-\alpha)Q_t(s,m^i(\tau))+\alpha(r+\gamma Q_t(s',m^i(\tau-1))).
\end{equation} 
}
As usual, we define 
 {
\begin{equation}
    \Delta_t(s,m^i(\tau))=Q_t(s,m^i(\tau))-Q^*(s,m^i(\tau)),
\end{equation}
}
 {
\begin{equation}
    F_t(s,m^i(\tau))=r+\gamma Q_t(s',m^i(\tau-1))-Q^*(s,m^i(\tau)).
\end{equation}
}
By subtracting $Q^*(s,m^i(\tau))$ from both sides of the learning rule, we again obtain an equation with the same form as the process in Lemma \ref{lemma:2}. Now we examine the expectation of $F_t(s,m^i(\tau))$,
\begin{equation} \label{equ:exp_multistep}
\begin{split}
    &\mathbb{E}[F_t(s,m^i(\tau))|P_n]\\
    =&\sum_{s'}P(s'|s,a)[r+\gamma Q_t(s',m^i(\tau-1))-Q^*(s,m^i(\tau))]\\
    =&(\boldsymbol{H}Q_t)(s,m^i(\tau))-Q^*(s,m^i(\tau))\\
    =&(\boldsymbol{H}Q_t)(s,m^i(\tau))-(\boldsymbol{H}Q^*)(s,m^i(\tau))\\
    \leq& \gamma \|Q_t-Q^*\|_\infty\\
    =&\gamma \|\Delta_t\|_\infty.
\end{split}
\end{equation}
For the same reason as (\ref{equ:var_onestep}), the variance of $F_t(s,m^i(\tau))$ is bounded,
\begin{equation} \label{equ:var_multistep}
    {\rm Var}[F_t(s,m^i(\tau))|P_n] \leq C(1+\| \Delta_t \| ^2_W).
\end{equation}

It can be observed from (\ref{equ:exp_onestep}) and (\ref{equ:exp_multistep}) that no matter which value $\tau$ takes $\mathbb{E}[F_t(s,m^i(\tau)|P_n] \leq \gamma \|\Delta_t\|_\infty$ always holds. The fact that the variance of any action is bounded is proved by (\ref{equ:var_onestep}) and (\ref{equ:var_multistep}). In conclusion, the conditions of Lemma \ref{lemma:2} are all satisfied, from which the convergence of EASpace is proved.
\end{proof}

\section{Results}\label{sec:experiments_and_results}
The proposed algorithm EASpace is firstly verified in the grid-based navigation domain used by \cite{fernandez2006probabilistic} and \cite{li2019context}. Then a multi-agent pursuit problem that has a larger continuous state space is considered. To demonstrate the efficiency of EASpace, several baseline algorithms are trained as follows.
\begin{itemize}
       \item DQN. This algorithm is proposed by \cite{silver2016mastering}, learning the target task from scratch.
        \item Shaping. This algorithm is proposed by \cite{brys2015reinforcement}, adding an additional reward to the actions demonstrated by expert policies. The potential of state-action pairs that are not demonstrated by expert policies is set to 0, while the potential of demonstrated state-action pairs is $p_{Shaping}$. 
        \item CAPS. This algorithm is proposed by \cite{li2019context}, employing the option-critic architecture \cite{bacon2017option} to learn when and which expert policy is best to reuse and when it terminates. To alleviate the option-shrink issue, the regularizer $\rho$ is introduced \cite{bacon2017option}. We construct a two-stream neural network as \cite{li2019context} does. The first stream outputs action values while the second stream outputs the corresponding termination probability. Note that the primitive actions will not be extended in our implementation because simply extending primitive actions is not promising in most complex problems. 
\end{itemize} 

Similar to EASpace, all the above benchmark algorithms are based on DQN. Therefore, we adjust the hyperparameters of DQN firstly. They are then copied to all other algorithms, including EASpace, due to the limitation of computing resources. The unique hyperparameters, \emph{e.g.} $\tau_0$ for EASpace and $\rho$ for CAPS, are fine-tuned via grid search for every algorithm. All the general and unique hyperparameters are listed in TABLE \ref{tab:hyperparamter}. The main contribution of PTF \cite{yang2020efficient} is orthogonal to EASpace, so it is not included as a baseline. Besides the aforementioned algorithms, ablation studies are performed as follows.

\begin{itemize}
       \item SMDP. IMALR is removed. The action value function is trained by SMDP {Q-learning}.
        \item No-Bonus. The macro action bonus is removed from EASpace. The agent is trained to maximize the task reward. 
\end{itemize} 

The code used in this paper and the video are available at https://github.com/Zero8319/EASpace.

\begin{table}
    \centering
    \caption{List of hyperparameters and their values}
    \label{tab:hyperparamter}
    \begin{tabular}{c|cc}
        \toprule
         & Hyperparameters & Values (grid/pursuit)\\
        \hline
        \multirow{10}{*}{General} & Learning rate $\alpha$ & 7e-5\\
         & $\gamma$ & 0.99\\
         & Maximal episode & 8000\\
         & Minibatch size & 128\\
         & Memory size & 1e6\\
         & Exploration strategy & $\epsilon$-greedy \\
         & Decay strategy of $\epsilon$ & Linear \\
         & Final exploration episode & 4000 \\
         & $f$ & 500\\
         & $T_u$ & 300 / 1000\\
         & $T$ & 300 / 1000\\
         \hline
         \multirow{2}{*}{EASpace} & $\tau_0$ & 10 / 20 \\
         & $c$ & 0.01 \\
         \hline
         Shaping & $p_{Shaping}$ & -0.05 / -0.5 \\
         \hline
         CAPS & $\rho$ & 0.01 / 0.1 \\
         \bottomrule
    \end{tabular}
\end{table}

\subsection{Grid-based Navigation} \label{subsec:grid_world}
\begin{figure}
    \centering
    \includegraphics[width=0.38\textwidth]{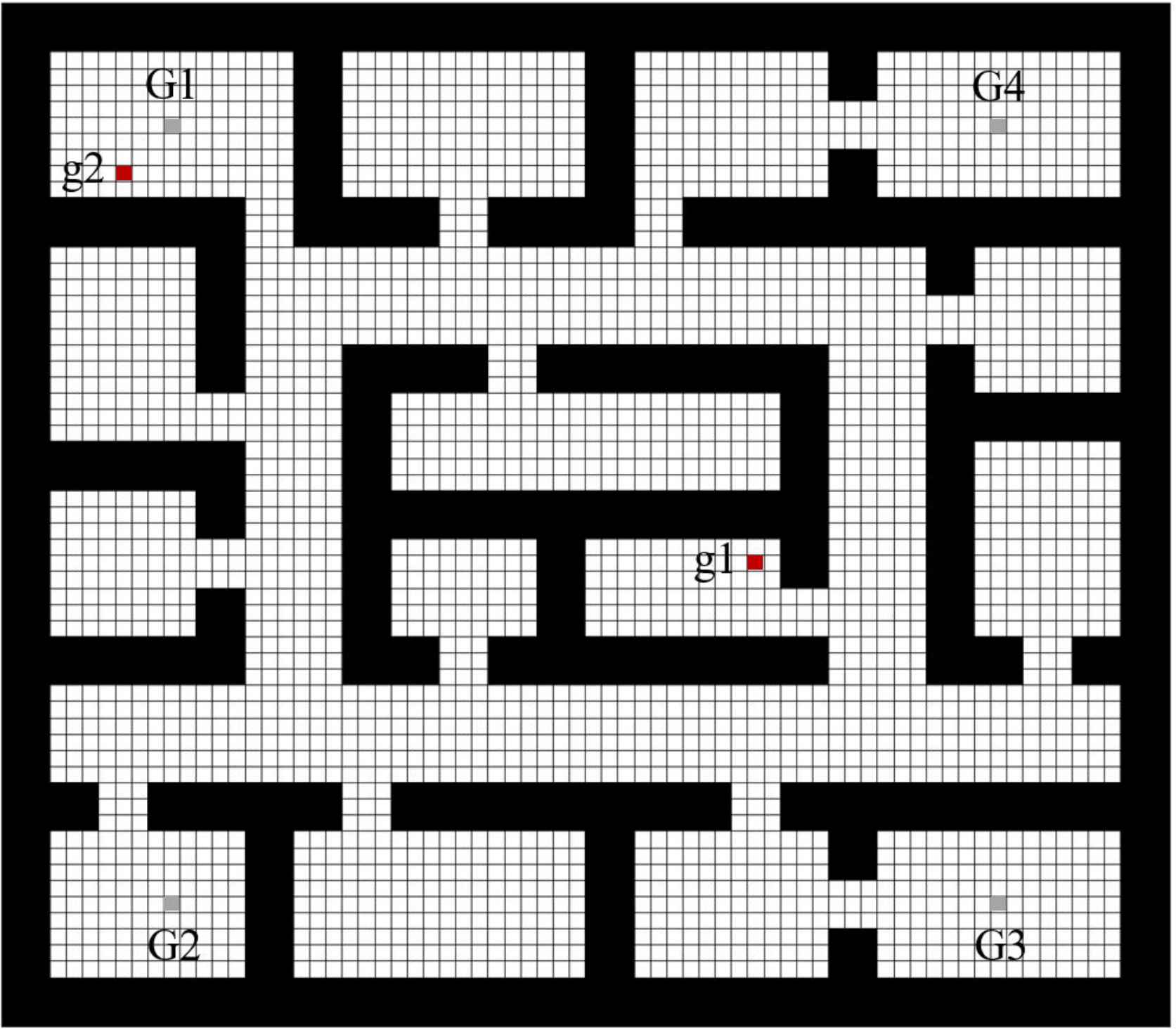}
    \caption{The map used in the grid-based navigation problems. G1, G2, G3, and G4 are goals for source tasks while g1, g2 are goals for target tasks.}
    \label{fig:maze}
\end{figure}

To create a long-horizon task, we enlarge the original maze in \cite{fernandez2006probabilistic,li2019context} by 3 times. As shown in Fig. \ref{fig:maze}, G1, G2, G3, and G4 denote the goals of four source tasks while g1 and g2 represent the goals of two target tasks. The state space is the $x-y$ coordinates in the maze. The action space is composed of $up, down, left, right$, each of them moving the agent to the desired adjacent grid with {a} probability of 0.8. To accelerate the learning process, besides the main reward that is 10 when the agent reaches the goal and 0 otherwise, an additional potential-based reward \cite{ng1999policy} is granted according to the Manhattan distance from the goal. The agent is randomly initiated in each episode. The action value function is approximated by a neural network that uses ReLU nonlinearities as activation functions besides the output layer. The neural network has 3 fully-connected layers with 64 units. Since the enlarged maze is a quite challenging task to learn from scratch, we train the source policies in the original small maze via DQN. A linear mapping is used to find the most similar state when they are employed as expert policies for target tasks. 

 \begin{figure}
    \centering
    \subfloat[task g1]{
    \includegraphics[width=0.48\textwidth]{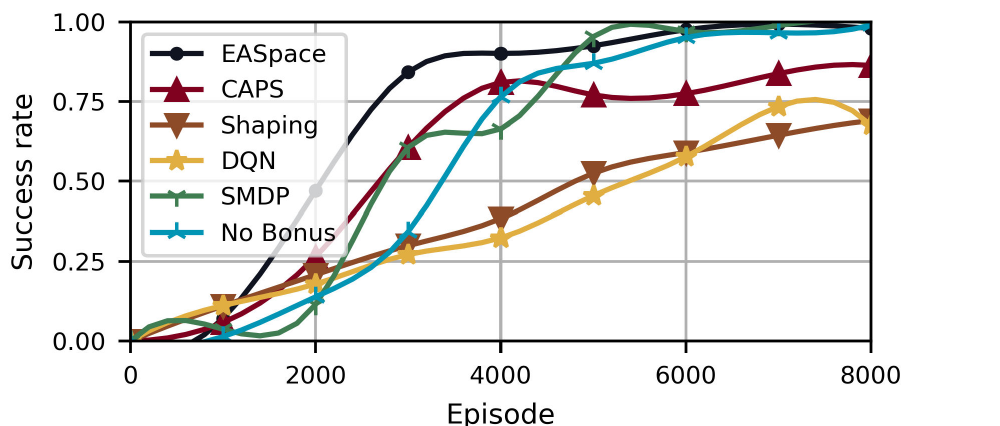}}
    \quad
    \subfloat[task g2]{
    \includegraphics[width=0.48\textwidth]{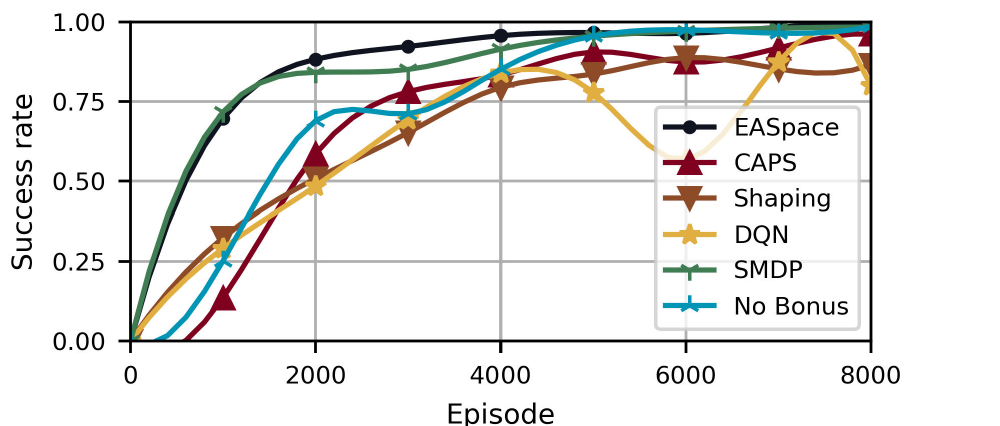}}
    \caption{The learning curves for the grid-based navigation problems. The success rate is averaged over 1000 validation episodes and 3 random seeds.}
    \label{fig:success_rate_maze}
\end{figure}

 The learning curves are demonstrated in Fig. \ref{fig:success_rate_maze}, while the data efficiency results are listed in TABLE \ref{tab:g1} and TABLE \ref{tab:g2} (No-Bonus is abbreviated to N.B.). As expected, DQN has the poorest performance as there is no external knowledge. Although Shaping transfers the expert knowledge by additional rewards, the long-horizon issue makes it difficult to distinguish useless advice from helpful ones at all timesteps. As a result, the agent usually follows the sub-optimal expert policies in the training process, thus making negative transfer inevitable. In comparison, CAPS employs macro actions to temporally abstract the learning process, which shortens the task horizon dramatically. Since credit assignment is much easier in such short-horizon tasks, useful macro actions can be determined efficiently. By learning from useful expert policies and abandoning useless ones, CAPS outperforms DQN and Shaping by a large margin. However, as shown in TABLE \ref{tab:g1} and TABLE \ref{tab:g2}, CAPS is inferior to EASpace in terms of both data efficiency and asymptotic performance. The reason is two-fold. First, EASpace provides a suitable exploration strategy for the learning of the optimal length of macro actions. For example, EASpace can employ long-duration macro actions to explore the environment from the very beginning of the training process. In comparison, CAPS needs to learn which macro action is better firstly, then extend it. So long-duration macro actions are available only after a lengthy learning process. Second, the macro action bonus enables EASpace to frequently exploit useful long-duration macro actions, which is significant to employ temporal abstraction for structured exploration and efficient credit assignment. In comparison, the action values of long-duration macro actions in CAPS are always lower than that of the best primitive action, making the useful macro actions rarely selected. TABLE \ref{tab:g1} and TABLE \ref{tab:g2} also indicate that EASpace outperforms SMDP and No-Bonus at most checkpoints, especially in the first 4000 episodes. As mentioned above, IMALR enables EASpace to extract more transitions in one episode, while the macro action bonus results in more successful experiences by executing useful long-duration macro actions. Both of them contribute to better performance in the early training process.

\begin{table}
    \centering
    \caption{The data efficiency results for task $g1$}
    \begin{tabular}{ccccccc}
        \toprule
        \label{tab:g1}
        Episodes & EASpace & CAPS & Shaping & DQN & SMDP & N.B.\\
        \midrule
        1000 & 0.07 & 0.06 & \textbf{0.11} & \textbf{0.11} & 0.04 & 0.01\\
        2000 & \textbf{0.47} & 0.26 & 0.21 & 0.18 & 0.11 & 0.14\\
        3000 & \textbf{0.84} & 0.61 & 0.30 & 0.27 & 0.61 & 0.34\\
        4000 & \textbf{0.90} & 0.81 & 0.38 & 0.32 & 0.66 & 0.77\\
        5000 & 0.93 & 0.77 & 0.53 & 0.46 & \textbf{0.95} & 0.54\\
        6000 & \textbf{0.98} & 0.77 & 0.59 & 0.58 & 0.97 & 0.95\\
        7000 & \textbf{0.99} & 0.84 & 0.66 & 0.73 & \textbf{0.99} & 0.97\\
        8000 & \textbf{0.99} & 0.86 & 0.69 & 0.68 & \textbf{0.99} & \textbf{0.99}\\
        \bottomrule
    \end{tabular}
\end{table}

\begin{table}
    \centering
    \caption{The data efficiency results for task $g2$}
    \begin{tabular}{ccccccc}
        \toprule
        \label{tab:g2}
        Episodes & EASpace & CAPS & Shaping & DQN & SMDP & N.B.\\
        \midrule
        1000 & 0.70 & 0.14 & 0.32 & 0.29 & \textbf{0.72} & 0.25\\
        2000 & \textbf{0.88} & 0.56 & 0.51 & 0.48 & 0.84 & 0.69\\
        3000 & \textbf{0.92} & 0.78 & 0.65 & 0.69 & 0.85 & 0.71\\
        4000 & \textbf{0.96} & 0.83 & 0.79 & 0.84 & 0.91 & 0.85\\
        5000 & \textbf{0.97} & 0.91 & 0.84 & 0.78 & 0.96 & 0.96\\
        6000 & 0.96 & 0.87 & 0.89 & 0.57 & 0.97 & \textbf{0.98}\\
        7000 & \textbf{0.98} & 0.92 & 0.85 & 0.87 & \textbf{0.98} & 0.96\\
        8000 & \textbf{0.99} & 0.96 & 0.87 & 0.80 & 0.98 & \textbf{0.99}\\
        \bottomrule
    \end{tabular}
\end{table}

\begin{figure}
    \centering
    \subfloat[task g1]{
    \includegraphics[width=0.23\textwidth]{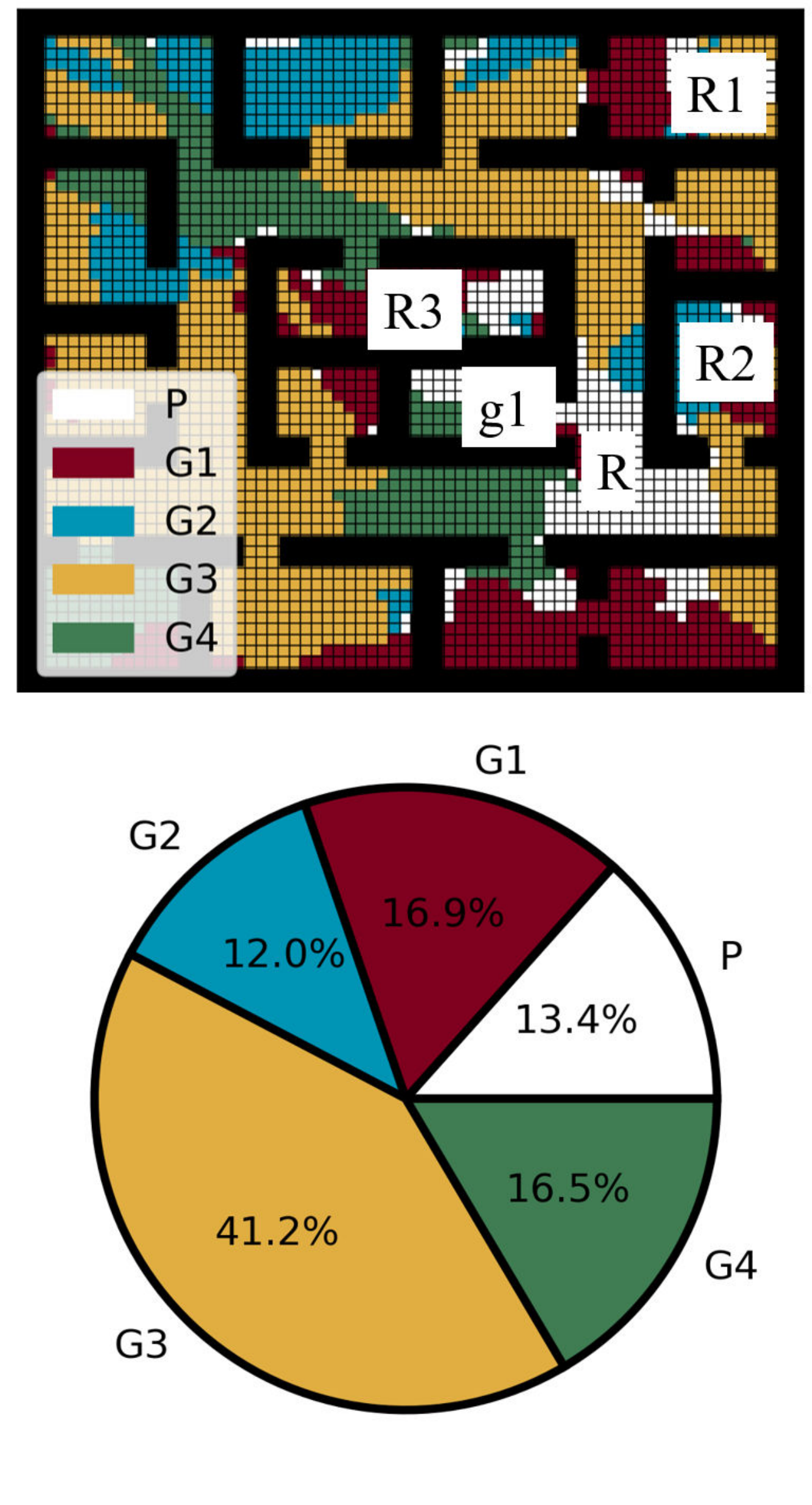}}
    \subfloat[task g2]{
    \includegraphics[width=0.23\textwidth]{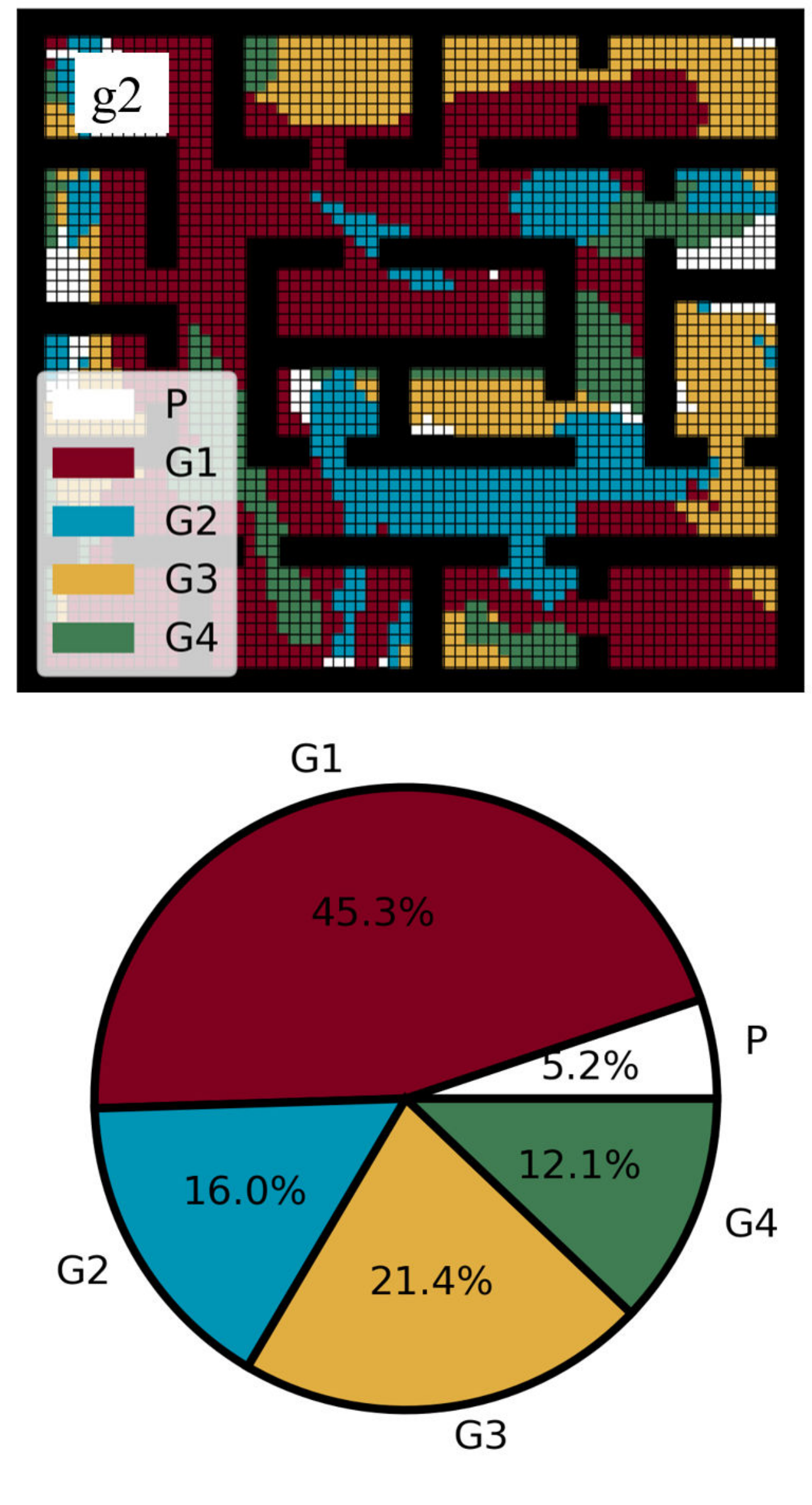}}
    \caption{The actions chosen by EASpace. The first row shows the distribution of actions with maximal action value. The action values are calculated for each grid, then the macro actions derived from the same expert policy are rendered with the same color regardless of their duration. The second row is the corresponding statistic results.}
    \label{fig:action_chosen}
\end{figure}

To demonstrate the learned policy by EASpace, the chosen action, \emph{i.e.} the action with maximal action value, is calculated for each grid, as illustrated in Fig. \ref{fig:action_chosen}. For the grid-based navigation problems, the main challenge is to get out of the rooms like R1, R2, and R3 in Fig. \ref{fig:action_chosen}(a) if the agent is initialized in them.  During the agent getting out of these rooms, it often receives negative rewards because it is walking away from the goal. However, structured exploration with expert policies guides the agent to the outside frequently, making it acquire more successful experiences. In the region near the goal, R in Fig. \ref{fig:action_chosen}(a), no expert policy guides the agent into the room where the goal is located. The agent abandons all expert policies, using primitive actions to reach the goal. This phenomenon demonstrates EASpace can learn when to execute macro actions and when to explore the environment by itself. Furthermore, the macro actions derived from the same expert policy cluster as illustrated in Fig. \ref{fig:action_chosen}, justifying the heuristics injected into EASpace: if the expert policy is helpful at a given state, so it is at the adjacent states. Fig. \ref{fig:action_chosen}(b) demonstrates the chosen actions in task g2. As expected, the expert policy G1 is selected more frequently than other macro actions due to its similarity with the target task. However, there are some grids in which other macro actions are executed. The first reason is that the executed macro action may be as good as the expert policy G1 in these grids. The second reason is that the expert policy G1 may be sub-optimal in these grids. Note that the source policies are trained using DQN, so there is no guarantee of the optimality of expert policies. When the expert policy G1 can not guide the agent to the goal, the agent resorts to other macro actions to get through the temporary difficult time. This phenomenon demonstrates EASpace can learn which macro action is the best.

\begin{figure}
    \centering
    \subfloat[task g1]{
    \includegraphics[width=0.48\textwidth]{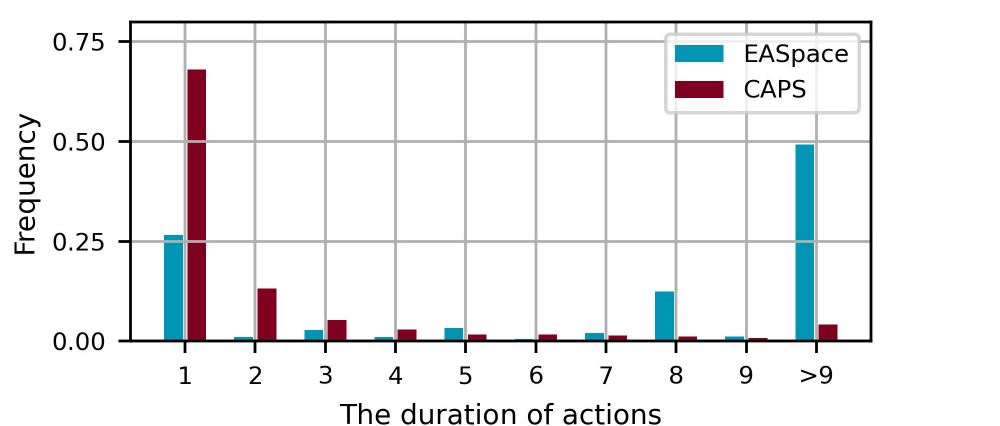}}
    \quad
    \subfloat[task g2]{
    \includegraphics[width=0.48\textwidth]{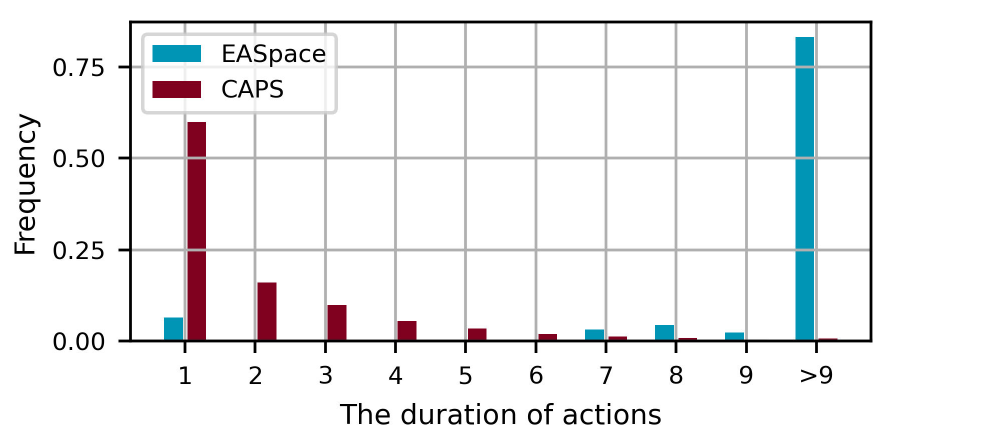}}
    \caption{The frequency of actions with different duration. The greater frequency means more time spent on the corresponding macro actions. The results are calculated over 1000 episodes. For EASpace the maximal duration of macro actions is 10 timesteps, so `$>9$' denotes 10-timestep macro actions. For CAPS, `$>9$' denotes all macro actions whose duration exceeds 9 timesteps.}
    \label{fig:action_length}
\end{figure}

To investigate the duration of macro actions, we perform 1000 validation episodes using the best policies learned by EASpace and CAPS. All the executed actions are recorded.  They are then categorized as illustrated in Fig. \ref{fig:action_length}. The frequency is defined as the duration divided by the total timesteps. For example, the frequency of 2-timestep macro actions is 0.1 means the agent executes 2-timestep macro actions for 10 timesteps in a 100-timestep episode. As shown in Fig. \ref{fig:action_length}, EASpace exploits long-duration macro actions more frequently, which suggests the significance of macro action bonus in the training process. Since the action value of long-duration macro actions is always less than that of short ones in theory, CAPS selects 1-timestep macro actions at most timesteps. The frequent decision-making processes not only lead to difficult credit assignment but also make CAPS fail to employ useful macro actions to acquire more successful experiences. Note that task g1 is not similar to any source task while the goal of task g2 is adjacent to G1. It is observed from Fig. \ref{fig:action_length} that EASpace efficiently learns the difference, increasing the execution frequency of long-duration macro actions dramatically in task g2. In comparison, CAPS fails to react to such a difference, employing similar frequency in both g1 and g2, which proves that EASpace is able to learn the optimal length of macro actions more efficiently.

\begin{figure}
    \centering
    \subfloat[the maximal length of macro actions $\tau_0$]{
    \includegraphics[width=0.48\textwidth]{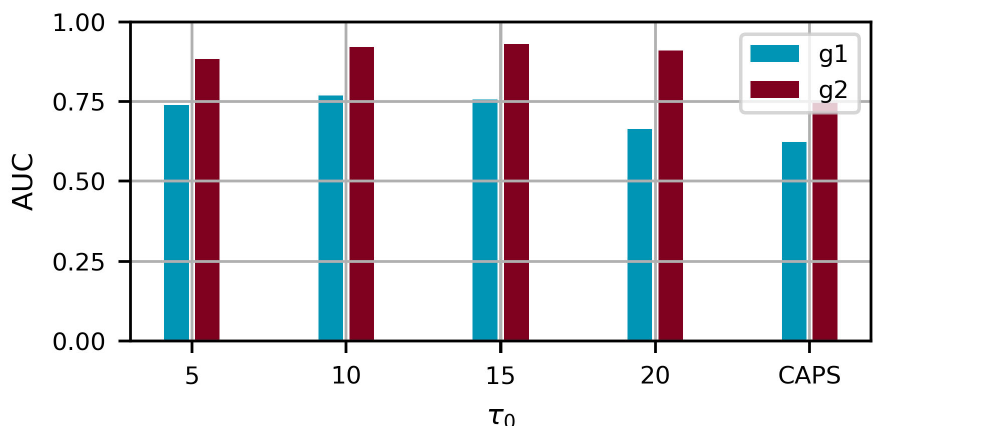}}
    \quad
    \subfloat[the scale factor of macro action bonus $c$]{
    \includegraphics[width=0.48\textwidth]{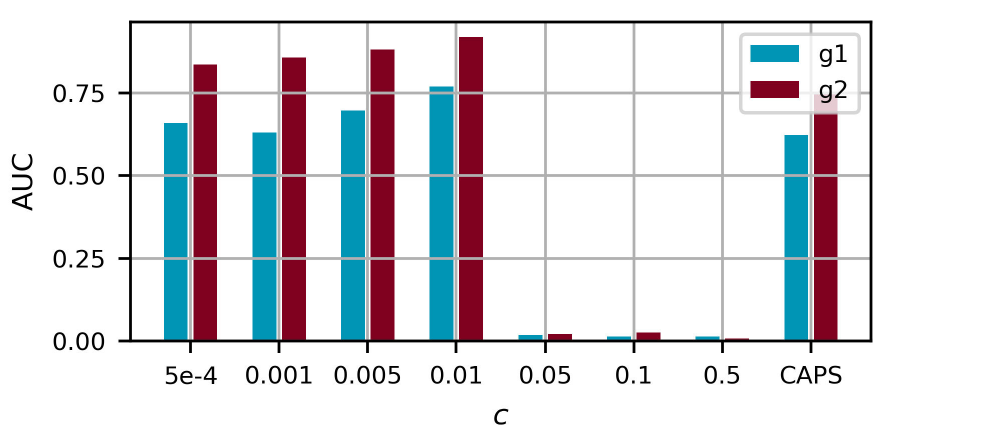}}
    \caption{The sensitivity of EASpace's unique hyperparameters in the grid-based navigation problems. The area under the learning curve (AUC) is normalized so that it is 1 for the agent that is capable to reach the goal from the beginning of the training process. The AUC of CAPS is also included for comparison.}
    \label{fig:sensitivity_maze}
\end{figure}

As shown in TABLE \ref{tab:hyperparamter}, the unique hyperparameters in EASpace are the maximal length of macro actions $\tau_0$ and the scale factor of macro action bonus $c$. We investigate their sensitivity in both task g1 and task g2. The performance of different hyperparameters is measured by the area under the learning curve (AUC). The corresponding results are demonstrated in Fig. \ref{fig:sensitivity_maze}.  As observed in Fig. \ref{fig:sensitivity_maze}(a), no matter what value $\tau_0$ takes, the resultant AUC is similar in task g2. Note that the size of the Q network is the same in all configurations while the size of the integrated action space differs dramatically. It proves that the integration of macro actions will not result in an overly large action space that hinders the learning process.  When $\tau_0=20$, the performance of EASpace in task g1 slightly degenerates as the dissimilarity between the target task and source tasks. It implies that overly long macro actions should be avoided if all expert policies are not suitable for the target task. However, although AUC is lower when $\tau_0$ takes sub-optimal values, EASpace still outperforms CAPS, which verifies the robustness of EASpace. Fig. \ref{fig:sensitivity_maze}(b) demonstrates the influence of different $c$. It is observed that if $c \leq 0.01$ the AUC of EASpace is much larger than CAPS while if $c > 0.01$ the policies learned by EASpace collapse. The reason is obvious. Since the macro action bonus is an intrinsic reward that will be added to external reward signals, only pursuing the intrinsic reward makes the agent forget the main task.

\subsection{Multi-agent Pursuit} \label{subsec:pursuit}

\begin{figure}
    \centering
    \subfloat[]{
    \includegraphics[width=0.15\textwidth]{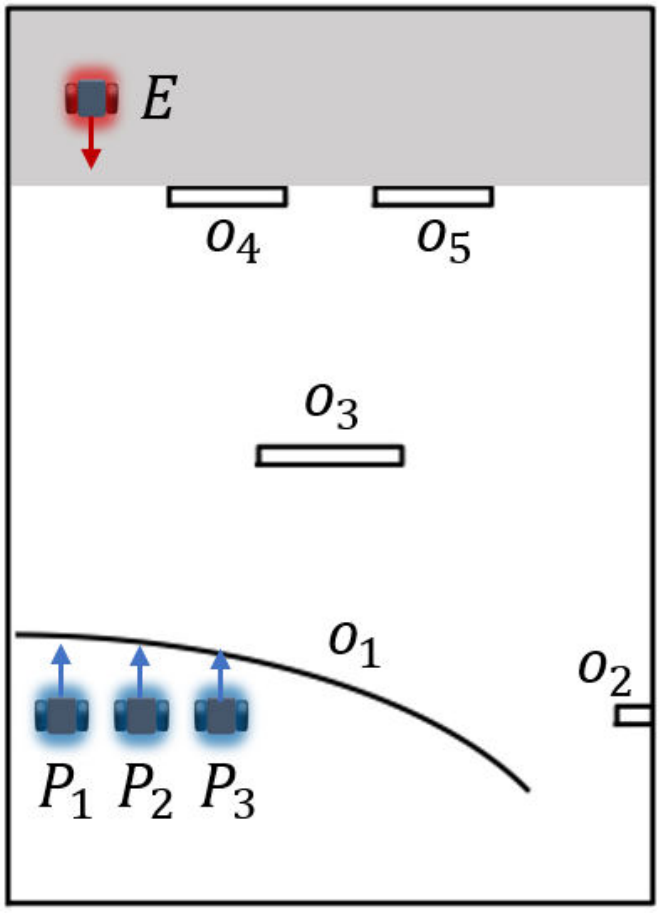}}
     \subfloat[]{
    \includegraphics[width=0.15\textwidth]{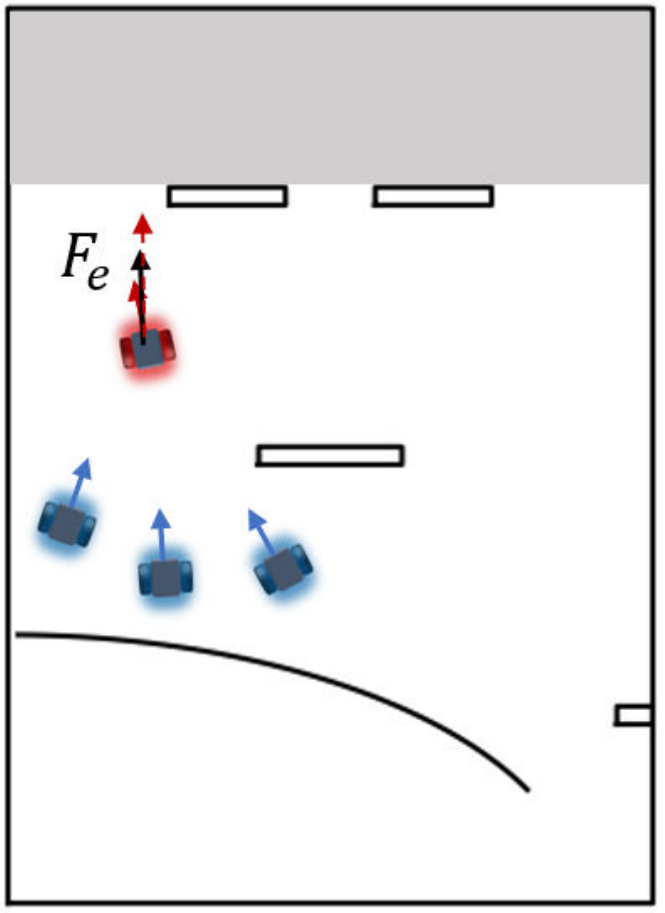}}
    \subfloat[]{
    \includegraphics[width=0.1535\textwidth]{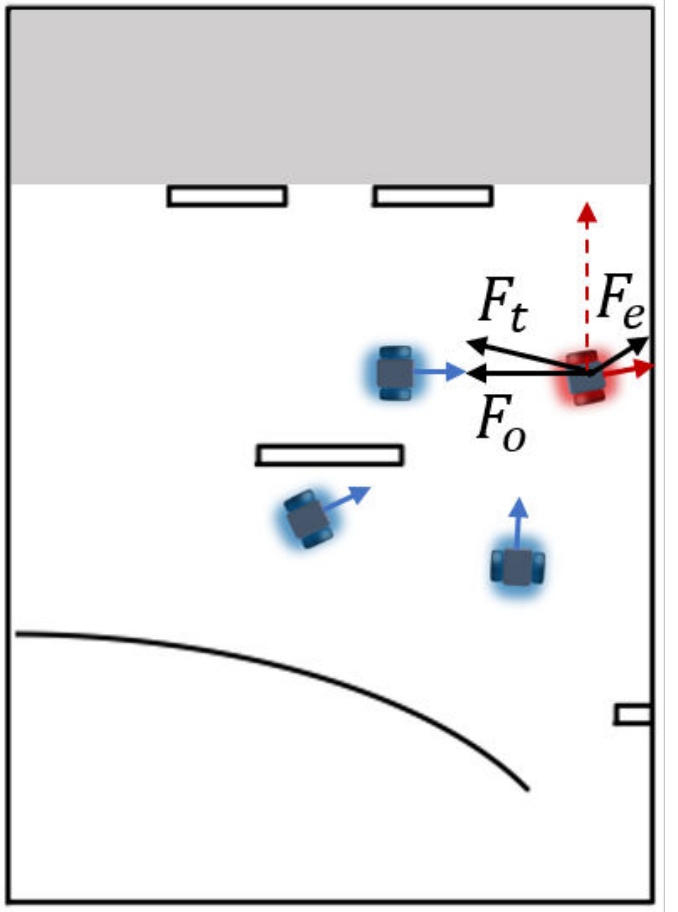}}
    \quad
    \subfloat[]{
    \includegraphics[width=0.15\textwidth]{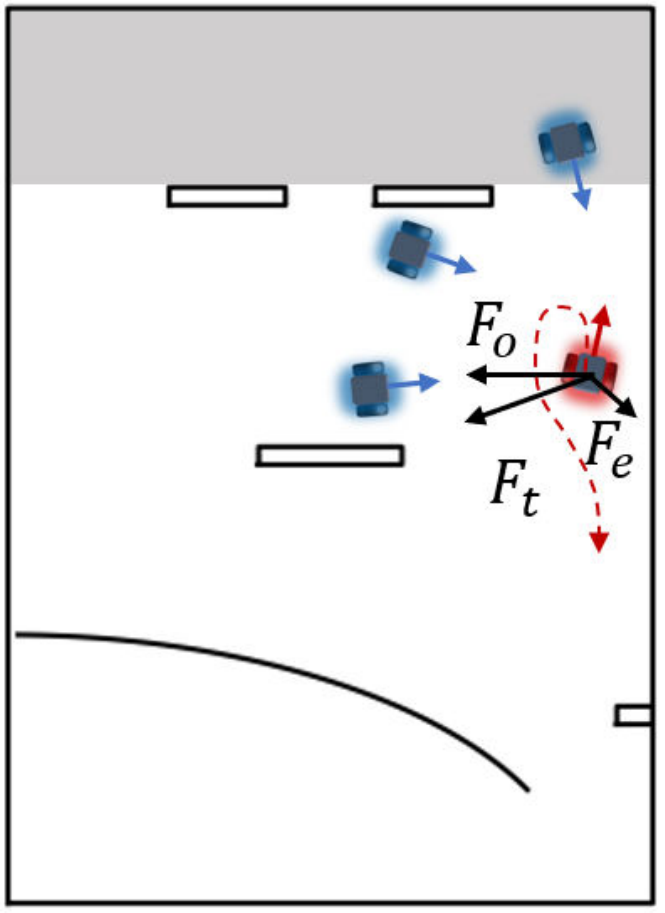}}
    \subfloat[]{
    \includegraphics[width=0.15\textwidth]{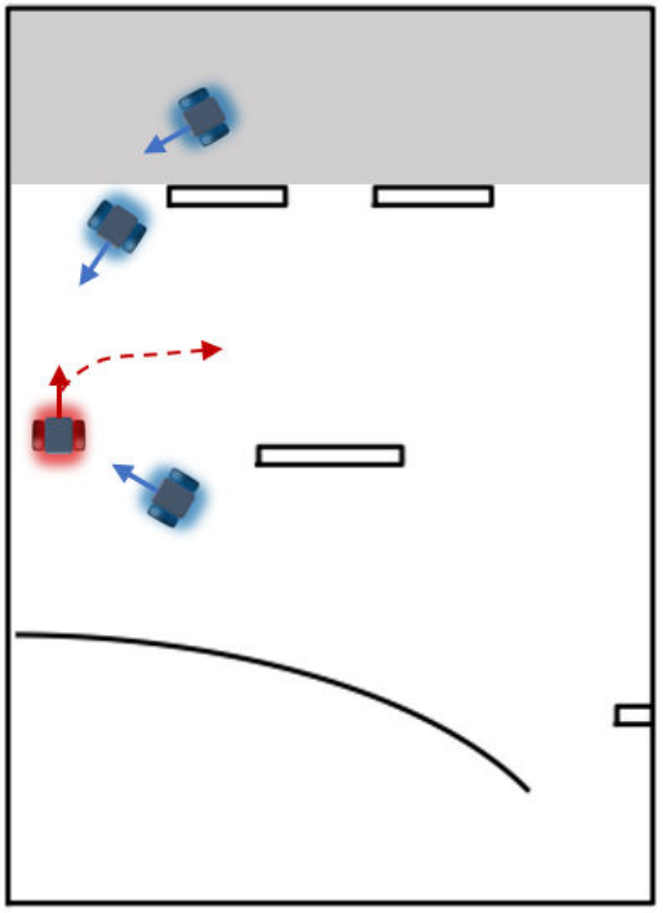}}
    \subfloat[]{
    \includegraphics[width=0.15\textwidth]{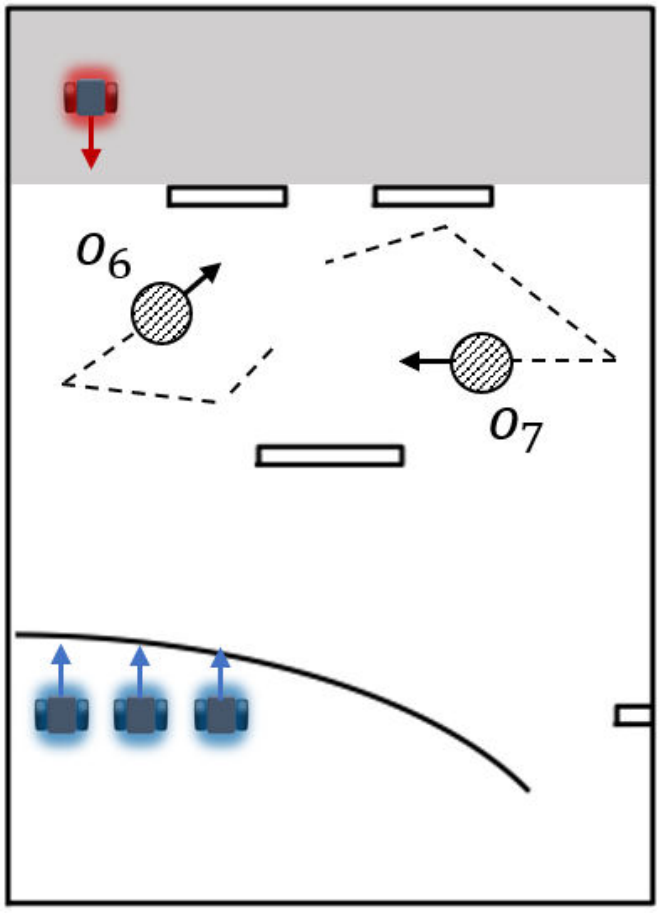}}
    \caption{(a) The initial state of the multi-agent pursuit problem. $E$ denotes the evader while $P_1,P_2,P_3$ denote three pursuers. $o_1,o_2,o_3,o_4,o_5$ are obstacles. The pursuit arena is confined. The evader is initialized in the grey region randomly at the beginning of each episode. (b)-(e) are typical escape strategies. The black arrows are the forces exerted on the evader. The red dotted line is the escaping trajectory of the evader. (b) Repulsed by pursuers. (c) Following the wall when the evader is in between pursuers and obstacles. (d) Turning around when there are pursuers in front of the evader. (e) Slipping through pursuers when the evader is encircled. {(f) The environment with two dynamic obstacles $o_6$ and $o_7$. The speed of dynamic obstacles is the same as that of pursuers. Dynamic obstacles periodically select a random direction and then move accordingly for $t_d \in [10,15]$ timesteps.}}
    \label{fig:escape}
\end{figure}

In this section, we consider a multi-agent pursuit problem that aims to coordinate three pursuers to capture one faster evader in a confined environment with obstacles. Fig. \ref{fig:escape}(a) illustrates the pursuit arena where $o_1,o_2,o_3,o_4,o_5$ denote obstacles. To capture the evader $E$, pursuers $P_1,P_2,P_3$ need to pass through the passage between $o_1$ and $o_2$ at first, then cooperate with each other to flank the faster evader. The speed of all pursuers and evader is assumed constant and the ratio is 3:4. The evader stops moving once any pursuer \emph{captures} it, \emph{i.e.} the distance between them is less than a threshold. The episode terminates when all pursuers capture the evader or the timesteps exceed 1000. The observations of pursuers include the relative distance and bearing of the evader, nearby obstacles, and nearby teammates. Hence, the environment is partially observable. The action space is 24 expected headings that uniformly discretize $360^{\circ}$. The reward function for each pursuer consists of four parts, $r=r_{main}+r_{time}+r_c+r_d$: (1) $r_{main}$, which is 50 when the pursuer captures the evader and 0 otherwise; (2) $r_{time}$, giving a penalty of -5 when the difference of headings between adjacent timesteps exceeds $45^{\circ}$; (3) $r_c$, giving a penalty of -50 when the pursuer collides with teammates or obstacles; (4) $r_d$, awarding the pursuer when it is approaching the evader according to the Euclidean distance while punishing it when moving away.

The escape policy derived from \cite{janosov2017group} is fixed during the learning process. It is a force-based method. At each timestep, the evader observes the positions of all pursuers, by which a repulsive force is calculated, $\boldsymbol{F}_{e}=\mathcal{N}[\sum_i(\boldsymbol{x}_e-\boldsymbol{x}_i)]$, where $\boldsymbol{x}_e$ is the position of the evader while $\boldsymbol{x}_i$ is the position of the pursuer. The operator $\mathcal{N}[\cdot]$ denotes the normalization of a vector. To avoid collisions, the evader is repulsed by nearby obstacles, $\boldsymbol{F}_r=\eta (\frac{1}{\| \boldsymbol{x}_e-\boldsymbol{x}_o \|}-\frac{1}{\rho_0}) \frac{1}{\| \boldsymbol{x}_e-\boldsymbol{x}_0 \|^2}\mathcal{N}[\boldsymbol{x}_e-\boldsymbol{x}_o]$, where $\boldsymbol{x}_o$ is the position of the nearest obstacle, $\eta$ is the scale factor, and $\rho_0$ is the influence range of obstacles. The total force $\boldsymbol{F}_t=\boldsymbol{F}_e+\boldsymbol{F}_o$ is employed to guide the evader. The wall following rules are introduced to smooth the trajectory of the evader near the obstacles \cite{9812083}. If the angle between $\boldsymbol{F}_t$ and $\boldsymbol{F}_e$ exceeds $90^{\circ}$, \emph{i.e.} the evader is in between pursuers and obstacles, it will move along the direction perpendicular to $\boldsymbol{F}_o$. Since there are two directions satisfying the perpendicularity, the evader prioritizes the one forming a smaller angle with the current heading to move smoothly (Fig. \ref{fig:escape}(c)). When there are pursuers near the prioritized direction, the evader chooses the other to escape from the pursuers (Fig. \ref{fig:escape}(d)). As \cite{janosov2017group} does, a slip rule is introduced to help the evader slip through pursuers when it is encircled (Fig. \ref{fig:escape}(e)). Totally, the escape policy is summarized as $[\boldsymbol{F}_t \vee wall following] \vee slip$, where $\vee$ denotes \emph{or}.

The first expert policy developed for the multi-agent pursuit problem is the artificial potential field (APF) method that guides the pursuer $P_i$ to a static goal via the combination of multiple forces \cite{koren1991potential}. The attractive force, $\boldsymbol{F}_{a}=\mathcal{N}[\boldsymbol{x}_e-\boldsymbol{x}_i]$ where $\boldsymbol{x}_e$ and $\boldsymbol{x}_i$ are the positions of evader and pursuer, guides the pursuer $P_i$ moving toward the evader. The repulsive force, $\boldsymbol{F}_r=\eta (\frac{1}{\| \boldsymbol{x}_i-\boldsymbol{x}_o \|}-\frac{1}{\rho_0}) \frac{1}{\| \boldsymbol{x}_i-\boldsymbol{x}_o \|^2}\mathcal{N}[\boldsymbol{x}_i-\boldsymbol{x}_o]$, prevents the pursuer $P_i$ from collisions with obstacles. The inter-individual force, $\boldsymbol{F}_{ind}=\sum_j {(0.5-\lambda / \|\boldsymbol{x}_j-\boldsymbol{x}_i \|}) \mathcal{N}[\boldsymbol{x}_j-\boldsymbol{x}_i]$ where $\boldsymbol{x}_j$ is the position of $P_i$'s teammates  and $\lambda$ is a positive number, repulses $P_i$ away from $P_j$ when they are close to each other while attracts $P_i$ to $P_j$ when they are far away. The total force, $\boldsymbol{F}_t=\boldsymbol{F}_a+\boldsymbol{F}_r+\boldsymbol{F}_{ind}$, is employed as the expected heading for the pursuer $P_i$. Although APF is just a path planning algorithm that is not designed for the multi-agent pursuit problem, Zhang \emph{et al.} pointed out that an appropriate inter-individual repulsive force can stimulate the emergence of cooperative pursuit strategies, \emph{e.g.} flank and encirclement \cite{9812083}. Therefore, APF is introduced as the first expert policy. Since APF is notorious for its local minima issue \cite{ge2002dynamic}, the wall following rules are developed as the second expert policy, which drives pursuers moving along the direction perpendicular to $\boldsymbol{F}_r$ (the vector pointing from the nearest obstacle to the pursuer). For simplicity, the pursuers always go around the obstacle anticlockwise. From the above description, it is obvious that both expert policies are sub-optimal for the pursuit problem. APF is useless when pursuers are close to obstacles or the evader, while the wall following rules do not consider the evader at all. Even in the states where some expert policy is useful, it is far from optimal due to the inflexible formulation.

In the training process, the centralized training decentralized execution paradigm is adopted with parameter sharing \cite{gupta2017cooperative}. It implies the experiences collected by all agents are used to train a sharing action-value function. Although the improved algorithm, D3QN \cite{wang2016dueling}, is used to train the pursuit policy, we denote it as `DQN' for consistency. The first two layers of the network have 64 units, after which two streams of fully-connected networks are employed. The first stream has 32 units, then outputs the action value for each action. The second stream has the same size with a state-value output. All the activation functions are ReLU nonlinearities.

\begin{figure}
    \centering
    \includegraphics[width=0.49\textwidth]{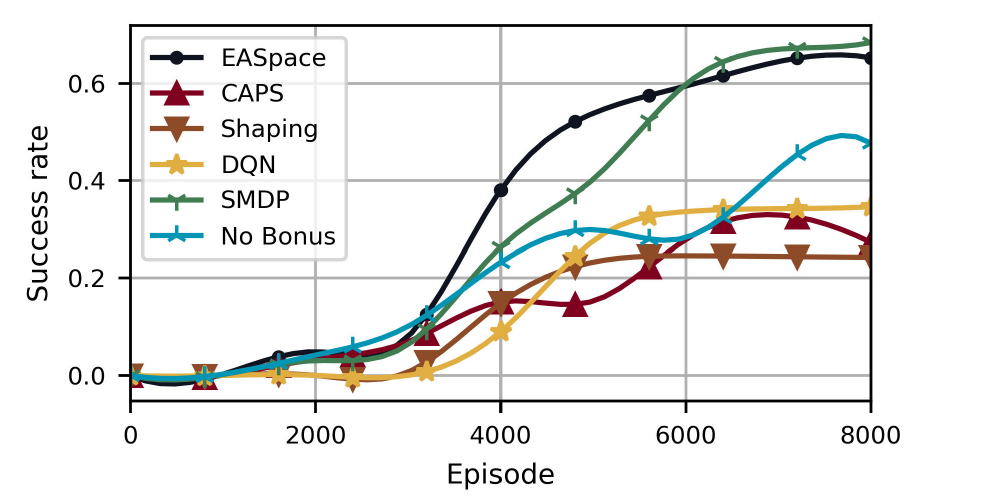}
    \caption{The learning curves for the multi-agent pursuit problem. The success rate is averaged over 1000 episodes and 3 random seeds.}
    \label{fig:pursuit}
\end{figure}

\begin{table}
    \centering
    \caption{The data efficiency results for the pursuit problem}
    \begin{tabular}{ccccccc}
        \toprule
        \label{tab:pursuit_table}
        Episodes & EASpace & CAPS & Shaping & DQN & SMDP & N.B.\\
        \midrule
        1000 & 0.00 & 0.00 & 0.00 & 0.00 & 0.00 & 0.00\\
        2000 & \textbf{0.05} & 0.03 & 0.00 & 0.00 & 0.03 & 0.04\\
        3000 & 0.08 & 0.07 & 0.00 & 0.00 & 0.06 & \textbf{0.10}\\
        4000 & \textbf{0.38} & 0.15 & 0.15 & 0.09 & 0.26 & 0.23\\
        5000 & \textbf{0.54} & 0.15 & 0.23 & 0.28 & 0.40 & 0.30\\
        6000 & \textbf{0.59} & 0.28 & 0.24 & 0.34 & \textbf{0.59} & 0.28\\
        7000 & 0.64 & 0.33 & 0.24 & 0.34 & \textbf{0.67} & 0.42\\
        8000 & 0.65 & 0.27 & 0.24 & 0.35 & \textbf{0.68} & 0.48\\
        \bottomrule
    \end{tabular}
\end{table}

Fig. \ref{fig:pursuit} and TABLE \ref{tab:pursuit_table} illustrate the success rate achieved by each algorithm (No-Bonus is abbreviated to N.B.). EASpace and SMDP outperform other algorithms in terms of both data efficiency and asymptotic performance. Since DQN and Shaping do not use macro actions to alleviate the long-horizon issue, their learning processes end up with poor performance. As mentioned above, CAPS firstly learns which expert policy is better, then reduce the termination probability of the high-action-value macro actions to extend their length. However, in the multi-agent pursuit problem, APF and wall following are neither optimal nor totally useless in most states. The received reward is indistinguishable when expert policies are executed for only one timestep. The usefulness of the better macro action is revealed only after long-term execution. Therefore, CAPS struggles to evaluate the difference between these two expert policies, which results in selecting APF near the obstacle $o_1$ while following the wall when the encirclement is needed. In comparison, the length of macro actions {does not} need to be extended in EASpace. Long-duration macro actions are accessible from the beginning of the training process. Successively selecting the same expert policy for multiple timesteps makes it easier to recognize the usefulness of expert policies. Therefore, EASpace is more efficient to exploit useful long-duration macro actions for structured exploration. Although SMDP achieves similar asymptotic performance as EASpace, its success rate is less from 3000 to 6000 episodes. This phenomenon verifies the significance of IMALR.

\begin{figure}
    \centering
    \subfloat[]{
    \includegraphics[width=0.15\textwidth]{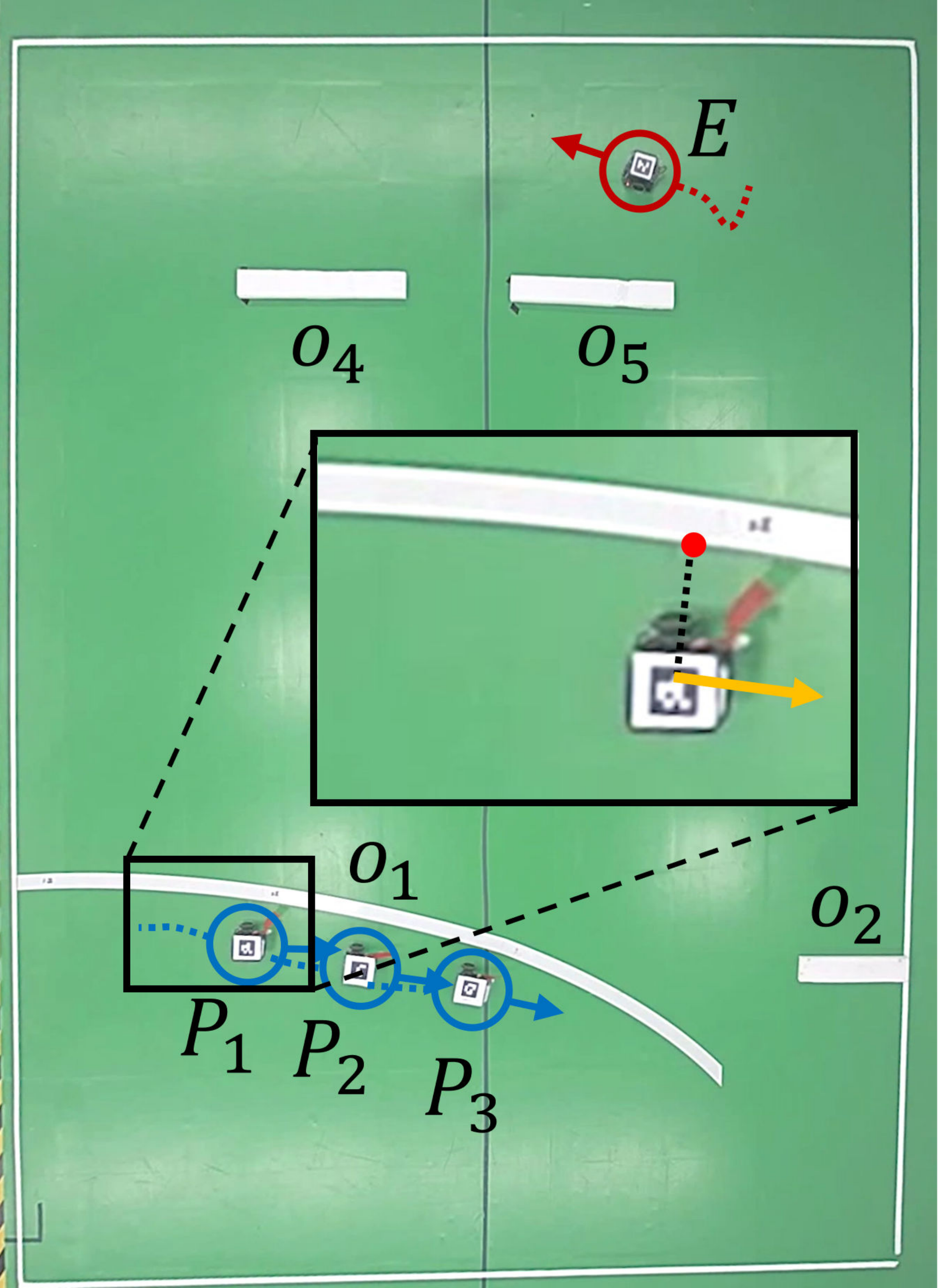}}
    \subfloat[]{
    \includegraphics[width=0.15\textwidth]{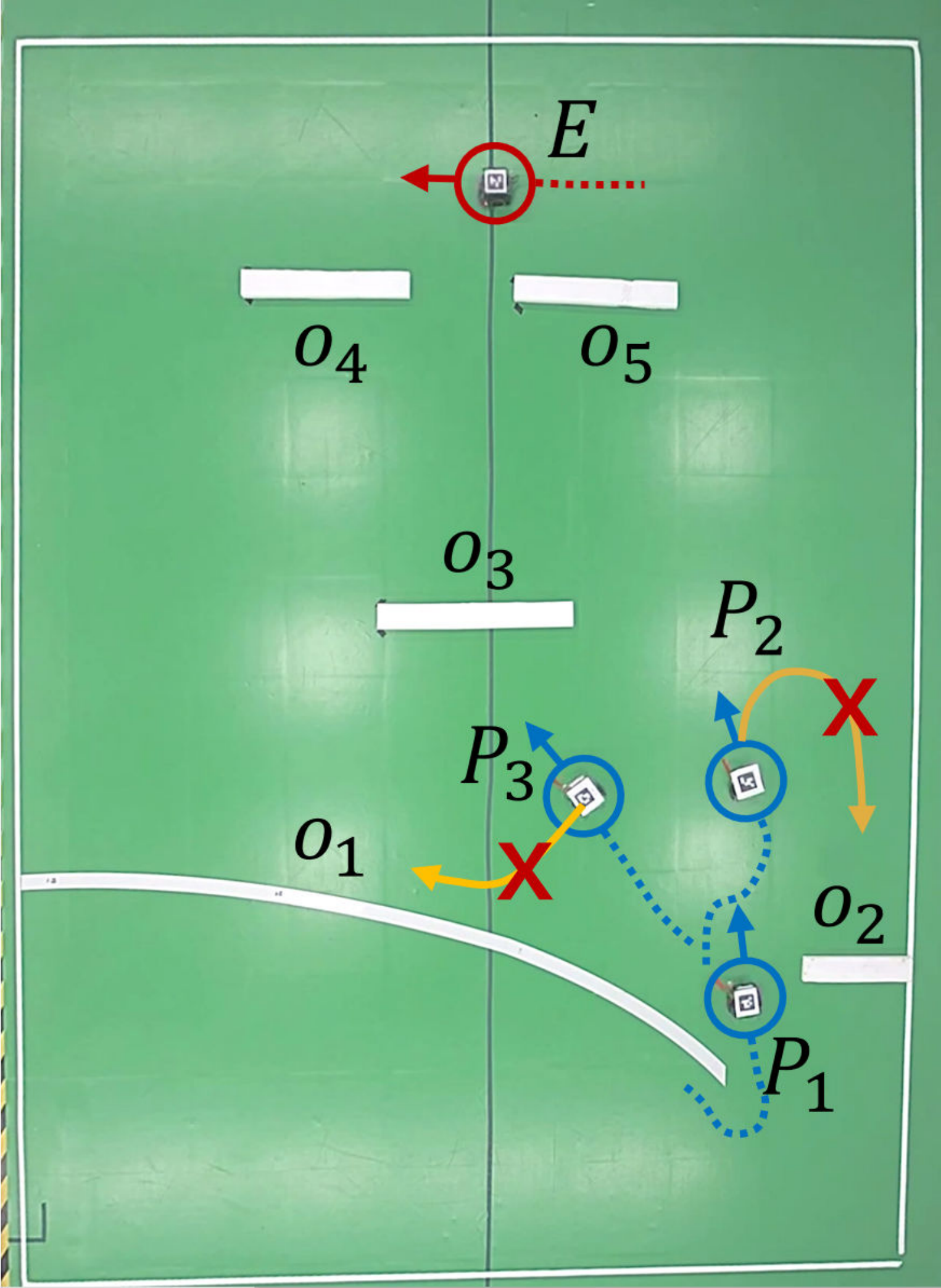}}
    \subfloat[]{
    \includegraphics[width=0.15\textwidth]{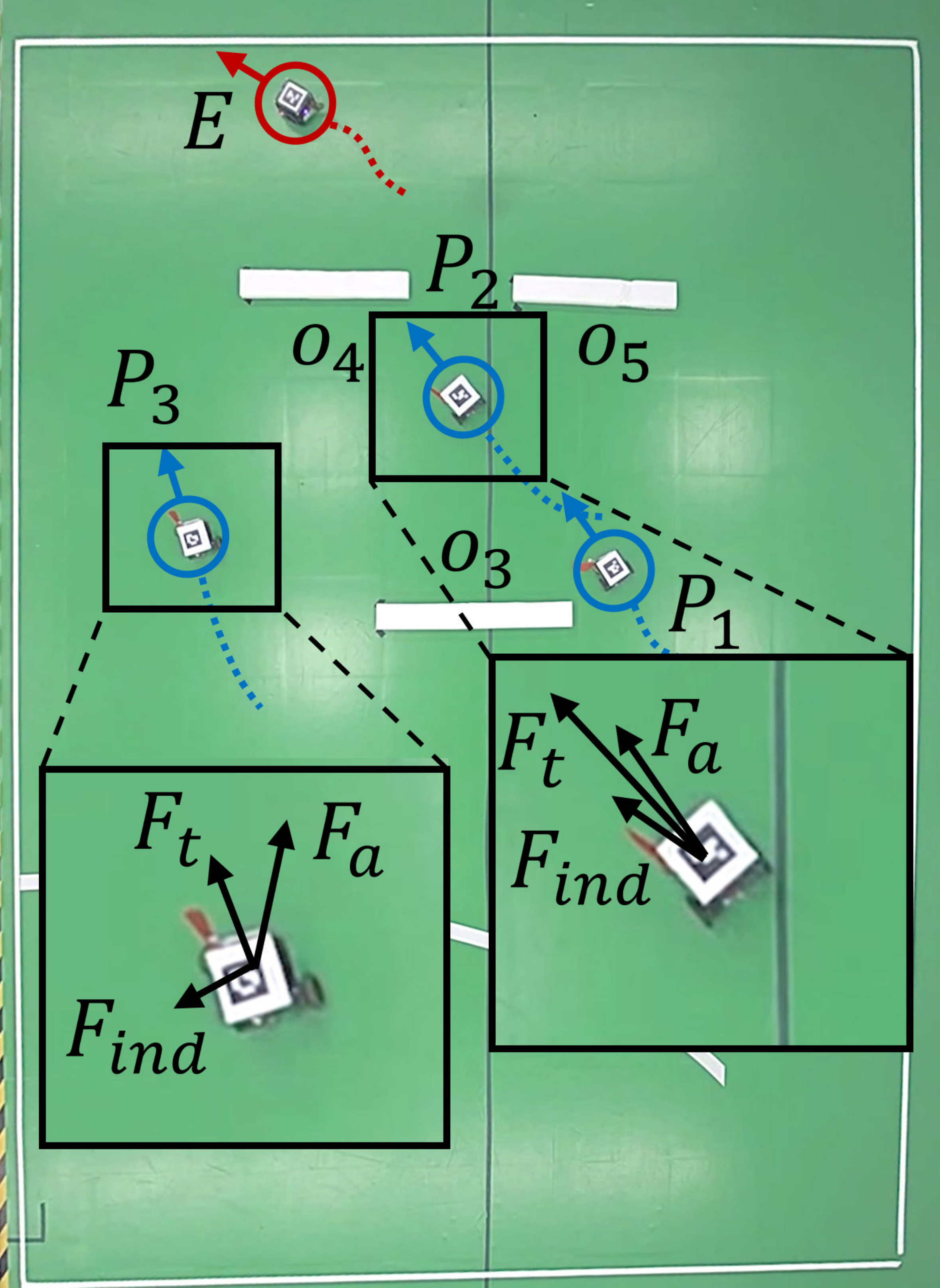}}
    \quad
    \subfloat[]{
    \includegraphics[width=0.15\textwidth]{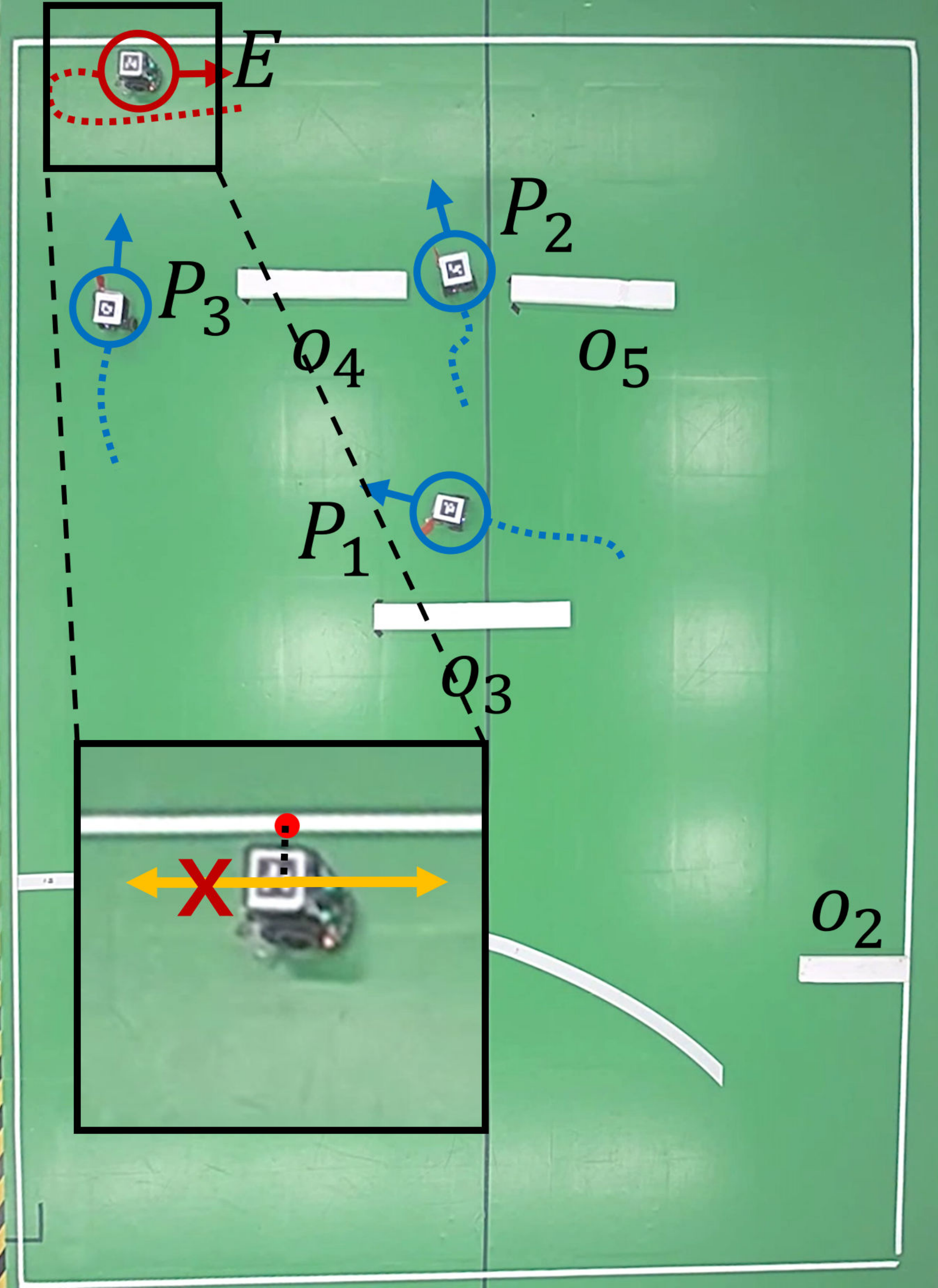}}
    \subfloat[]{
    \includegraphics[width=0.15\textwidth]{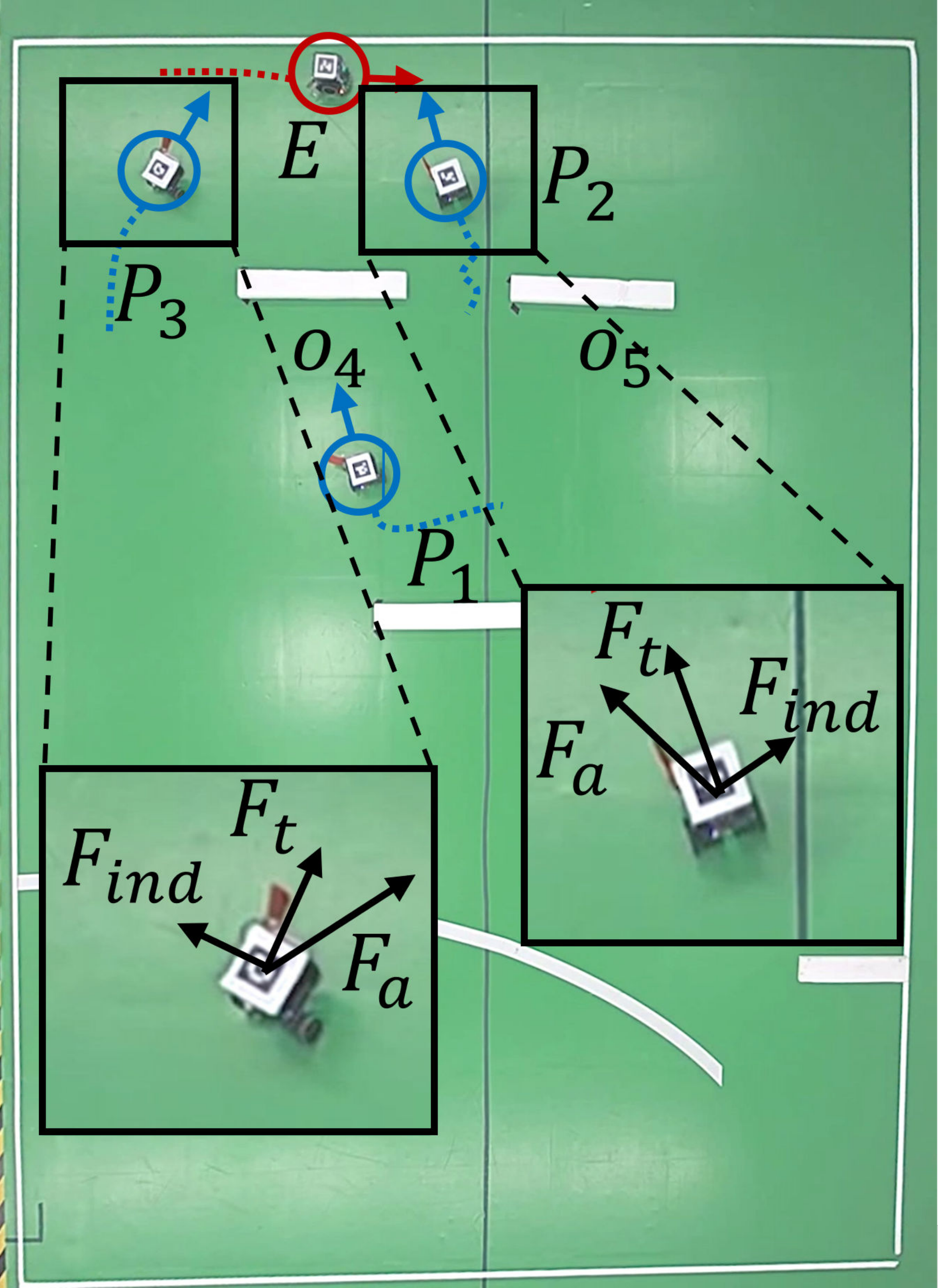}}
    \subfloat[]{
    \includegraphics[width=0.15\textwidth]{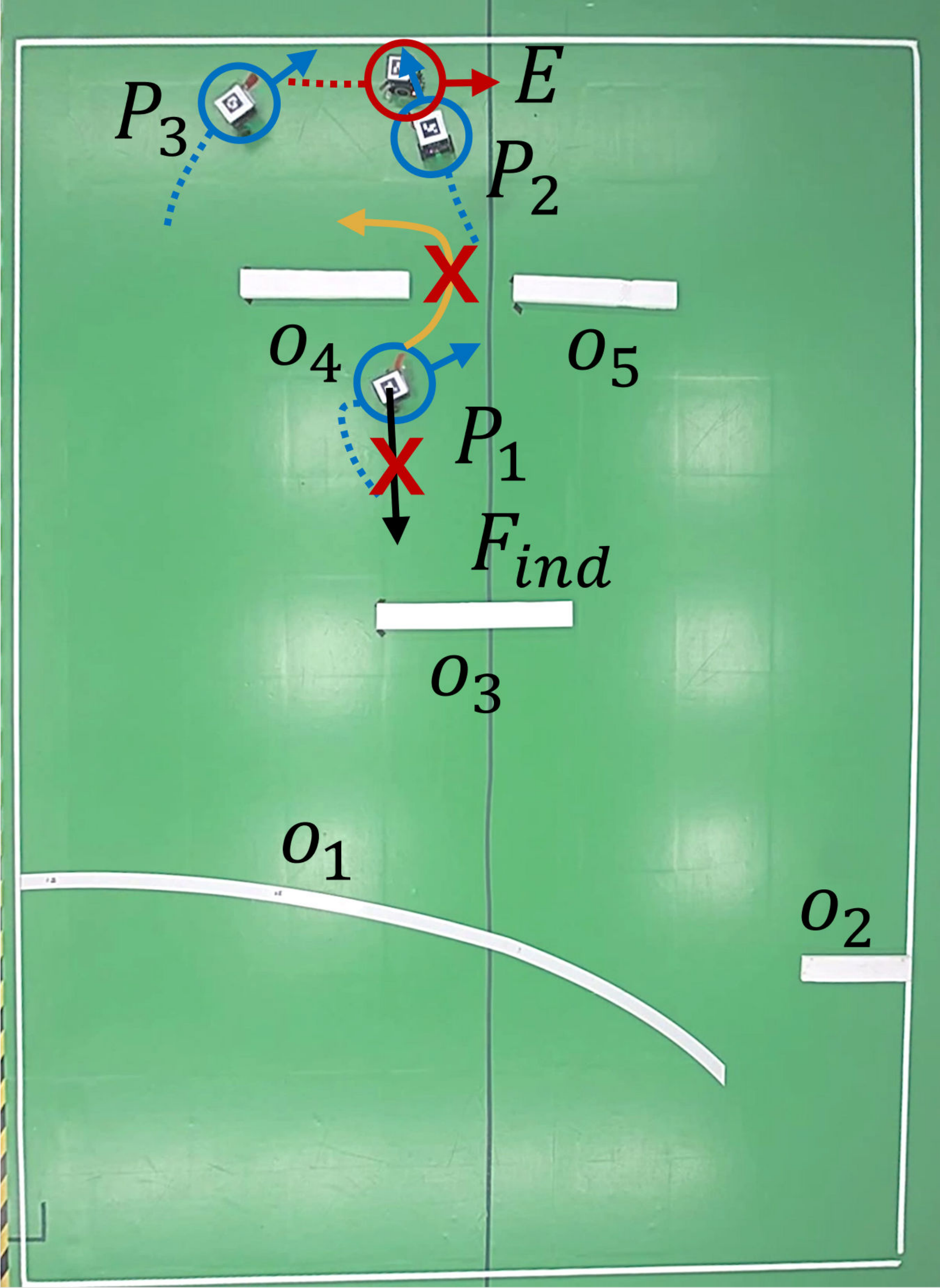}}
    \caption{The snapshots of the real-world experiment. The blue and red rings denote the pursuers $P_1,P_2,P_3$ and the evader $E$ respectively, while the attached arrows denote the current headings. The blue and red dotted lines are the history trajectories. $o_1,o_2,o_3,o_4,o_5$ are obstacles in the pursuit arena. The red circles in (a) and (d) are the nearest obstacle points. The yellow and black arrows in the overlapped plots denote the suggestions of wall following rules and artificial potential field (APF) methods, respectively. The red crosses on those suggestions imply that they are totally wrong advice.}
    \label{fig:pursuit_experiment}
\end{figure}

To demonstrate the learned policies of EASpace, the real-world experiment is conducted by directly deploying them to differential two-wheeled cars. For the descriptions of the experiment platform please refer to \cite{9812083}. We extract several key snapshots in one episode to qualitatively analyze the cooperative behaviors. The full video is available at https://github.com/Zero8319/EASpace. In Fig. \ref{fig:pursuit_experiment}(a), the pursuers $P_1,P_2,P_3$ execute the wall following rules to bypass the concave obstacle $o_1$. It implies that they move along the direction perpendicular to the line between the nearest obstacle (the red circle) and itself. Note that the wall following rules do not lead to the shortest trajectory. The pursuers learn to choose this sub-optimal but good enough macro action due to the macro action bonus. In Fig. \ref{fig:pursuit_experiment}(b), all pursuers go through the narrow passage between $o_1$ and $o_2$ one by one. After that, they abandon the wall following rules that lead them to move away from the evader meaninglessly, choosing APF or primitive actions to approach the evader.  In Fig. \ref{fig:pursuit_experiment}(c), the pursuers $P_1$ and $P_2$ resort to APF to keep an appropriate distance from each other. The repulsive inter-individual force $F_{ind}$ drives $P_3$ flanking from the left side, while $P_2$ chases directly from the gap between $o_4$ and $o_5$. The selection between APF and wall following demonstrates EASpace can learn which expert policy is useful at a certain state. In Fig. \ref{fig:pursuit_experiment}(d), the evader turns around because $P_3$ is approaching. However, the encirclement is complete. It should be highlighted that the headings of $P_2$ and $P_3$ do not point to the evader directly in Fig.\ref{fig:pursuit_experiment}(e). The individual force helps them `forecast' the evader's escape route though only the position of the evader can be observed.  In Fig. \ref{fig:pursuit_experiment}(f), the pursuer $P_2$ captures the evader. However, the episode ends when all pursuers are within the adjacent region of the evader. To accomplish the task the pursuer $P_1$ must resort to primitive actions because the individual force will repulse it away while wall following rules never guide it to the evader. The ultimate success proves EASpace can learn when to execute expert policies and when to abandon them.

\begin{figure}
    \centering
    \subfloat[the maximal length of macro actions $\tau_0$]{
    \includegraphics[width=0.48\textwidth]{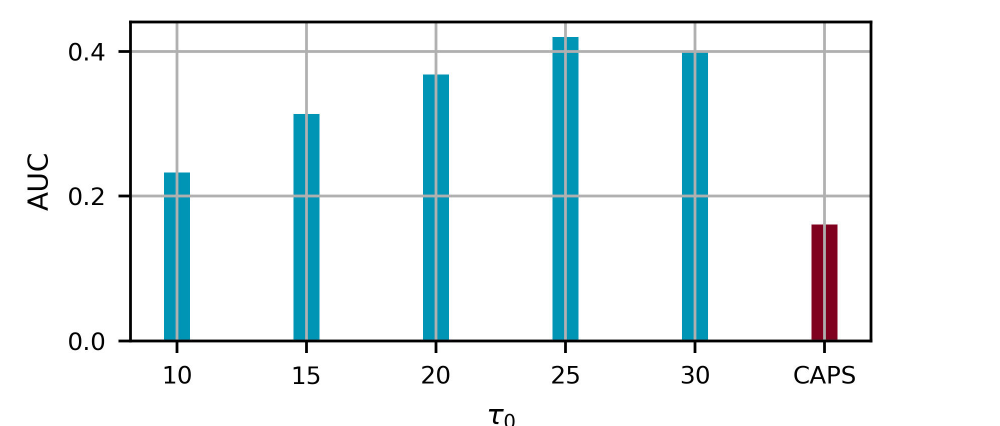}}
    \quad
    \subfloat[the scale factor of macro action bonus $c$]{
    \includegraphics[width=0.48\textwidth]{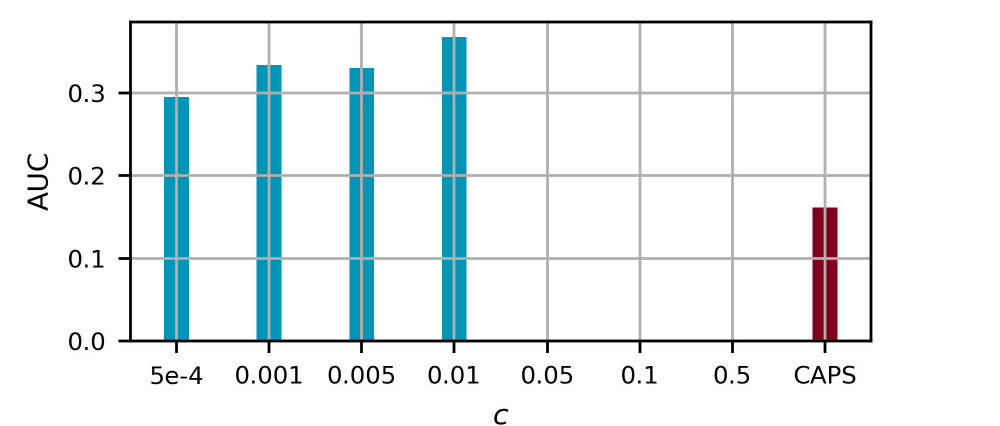}}
    \caption{The sensitivity of EASpace's unique hyperparameters in the multi-agent pursuit problems. The area under the learning curve (AUC) is normalized so that it is 1 for the agents that are capable to capture the evader from the beginning of the training process. The AUC of CAPS is also included for comparison. Note that the AUC is 0 when $c=0.05,0.1,0.5$.}
    \label{fig:sensitivity_pursuit}
\end{figure}

The sensitivity of hyperparameters in EASpace is also investigated for the pursuit problem. The results are shown in Fig. \ref{fig:sensitivity_pursuit}. As expected, the tendency of AUC is similar to that in the grid-based navigation problems. The only difference is that the small $\tau_0$ results in degenerated performance in the pursuit problem. The reason is that if the maximal length of macro actions $\tau_0$ is small, EASpace can not take full advantage of long-duration macro actions to structurally explore such a long-horizon task. The problems encountered by vanilla RL methods, \emph{e.g.} difficult credit assignment, also trouble EASpace. However, there is still a broad range of hyperparameters where EASpace outperforms baseline algorithms by a large margin, which justifies the robustness of EASpace.

{One potential concern over EASpace is its generalization ability. When the learned policies are directly deployed in the validation environment, EASpace suffers from the fact that the running optimal expert policy may turn sub-optimal during its execution due to the different state transition functions. If the agents keep executing this sub-optimal expert policy in the following timesteps, it is possible to result in a task failure. Therefore, the technique of \emph{interrupting macro actions} (IMA) is introduced from \cite{sutton1999between}. IMA empowers agents to early terminate the running macro action $m^{i_0}(\tau_0)$ when some other macro action $m^{i}(\tau)$ has larger values, \emph{i.e.}, $Q(s_0,m^{i}(\tau))>Q(s_0,m^{i_0}(\tau_0))+c_L$, where $c_L$ is a tunable threshold. To investigate the effectiveness of IMA, a validation environment is created by adding two dynamic obstacles into the training environment, as shown in Fig. \ref{fig:escape}(f). All the policies learned from Fig. \ref{fig:escape}(a) are tested directly without further training.} 

\begin{figure}
    \centering
    \subfloat[]{
    \includegraphics[width=0.48\textwidth]{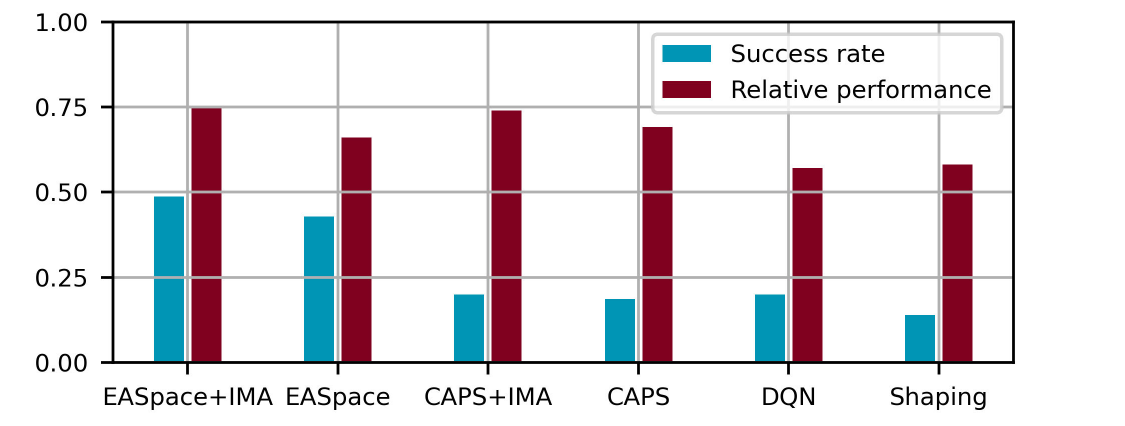}}
    \quad
    \subfloat[]{
    \includegraphics[width=0.48\textwidth]{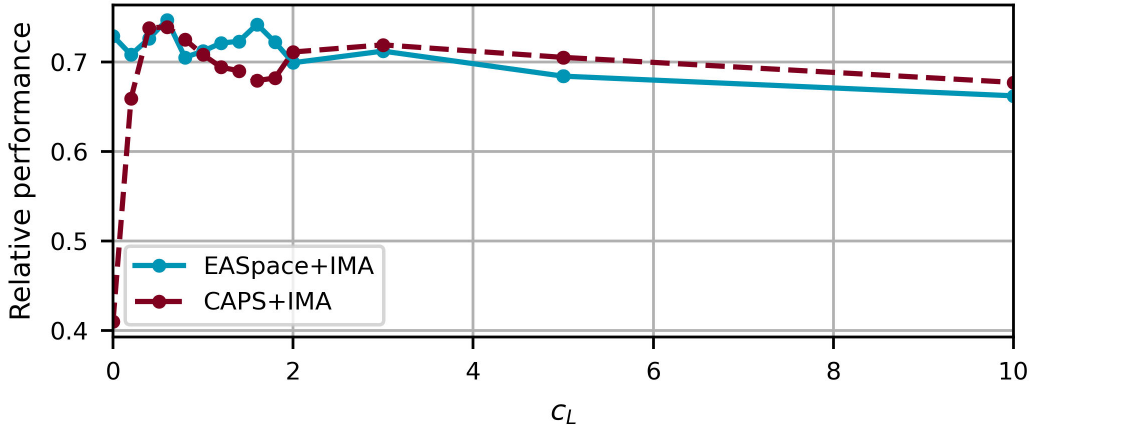}}
    \caption{{(a) The results of directly deploying the learned policies in the environment with two dynamic obstacles. The success rate is evaluated over 1000 episodes. The relative performance is defined as the ratio of the success rate in the validation settings (with dynamic obstacles) to that in the training settings (without dynamic obstacles). So larger relative performance indicates a greater ability to avoid dynamic obstacles. $c_L$ is manually tuned for both EASpace+IMA and CAPS+IMA. (b) The sensitivity of $c_L$.}}
    \label{fig:dynamic_obstacle}
\end{figure}

{The results shown in Fig. \ref{fig:dynamic_obstacle}(a) demonstrate that EASpace outperforms all benchmark algorithms in terms of success rate no matter whether it is integrated with IMA or not. It is also observed that the relative performance of EASpace, which is defined as the ratio of the success rate in the validation settings to that in the training settings, is slightly lower than that of CAPS. The reason is that CAPS evaluates the termination probability of macro actions at each timestep so that it has a chance to terminate sub-optimal expert policies once the environment states deviate from expectation. On the contrary, the sub-optimal expert policies terminate only when some predefined execution period is met in EASpace. However, as Fig. \ref{fig:dynamic_obstacle}(a) shows, EASpace+IMA achieves the best relative performance among all algorithms of interest, even better than the combination of CAPS and IMA. It proves that the aforementioned limitations of EASpace could be easily alleviated by the simple technique IMA so that it has a competitive generalization ability to deal with unseen environments. DQN and Shaping acquire the lowest relative performance, verifying that following expert policies is more promising to avoid dynamic obstacles. Fig. \ref{fig:dynamic_obstacle}(b) shows the sensitivity of $c_L$. When $c_L=0$, the current macro action is terminated as long as its action value is not the largest. Since the action values of macro actions are always no more than that of primitive actions in CAPS, the frequent termination of long-duration macro actions is inevitable for CAPS+IMA, which results in the dramatic performance deterioration as shown in Fig. \ref{fig:dynamic_obstacle}(b). In comparison, the macro action bonus enables EASpace+IMA to maintain satisfactory performance in a wide range of $c_L$, which further demonstrates the feasibility and efficiency of the proposed algorithm. }

\section{Conclusions and Perspectives}\label{sec:conclusion}
In this paper, a multi-policy transfer algorithm called EASpace was proposed to learn when to execute which expert policy and how long it lasts by enhancing the action space with macro actions. The special formulation of macro actions in EASpace provided a suitable exploration strategy that prompts the learning of the optimal length of macro actions. In addition, the explicit characterization of the length of macro actions makes it convenient to inject preference for long-duration macro actions, which is significant to exploit useful long-duration macro actions for efficient credit assignment and structured exploration. To extract more transitions, IMALR was proposed by adjusting the TD target of macro actions, after which the theoretical analysis was provided to guarantee its convergence and optimality. Experiments were conducted in the grid world and a multi-agent pursuit problem. The results consistently verified that EASpace outperforms multiple baseline algorithms in terms of data efficiency and asymptotic performance. The ablation study was designed by removing IMALR and macro action bonus from EASpace, from which we demonstrated that traditional SMDP methods are not efficient enough and the additional intrinsic reward is significant. The sensitivity test was conducted in both tasks, whose results verified the robustness of EASpace to sub-optimal hyperparameters. {The limitations of EASpace were also discussed in the multi-agent pursuit problems and the possible solutions were then provided. Simulation results proved that EASpace could acquire competitive generalization ability via slight modifications.} However, the {introduction} of the macro action bonus biases the original action value function. Therefore, in future works some advanced exploration techniques \cite{pathak2017curiosity,tutsoy2023novel} or curriculum learning methods \cite{narvekar2020curriculum} may be employed to keep the original reward function untouched. Besides, combining EASpace with model-based RL is another promising approach to further improve the data efficiency \cite{tutsoy2016analysis,tutsoy2016chaotic}.

\bibliography{references}
\bibliographystyle{IEEEtran}
\end{document}